\documentclass[12pt,preprint,epsf]{aastex}
\usepackage{natbib}

\slugcomment{ }
\shorttitle{SN\,2014J}
\shortauthors{ et al.}
\usepackage{epstopdf}
\usepackage{graphicx}
\usepackage{amsmath}
\usepackage{CJKutf8}

\begin{document}
\title{MAPPING CIRCUMSTELLAR MATTER WITH POLARIZED LIGHT -- THE CASE OF SUPERNOVA 2014J IN M82}
\shorttitle{Late-time SN\,2014J}
\author{Yi Yang\begin{CJK*}{UTF8}{gbsn}
(杨轶)
\end{CJK*}\altaffilmark{1,2}, 
        Lifan Wang\altaffilmark{1,3},
        Dietrich Baade\altaffilmark{4}, 
        Peter.~J. Brown\altaffilmark{1},
        Aleksandar Cikota\altaffilmark{4}, 
        Misty Cracraft\altaffilmark{5}, 
        Peter A. H\"oflich\altaffilmark{6}, 
        Justyn R. Maund\altaffilmark{7}{$^{,\dagger}$}, 
        Ferdinando Patat\altaffilmark{4},
        William~B. Sparks\altaffilmark{5}, 
        Jason Spyromilio\altaffilmark{4},
        Heloise F. Stevance\altaffilmark{7}, 
        Xiaofeng Wang\altaffilmark{8}, 
        J. Craig Wheeler\altaffilmark{9}}

\altaffiltext{1}{George P. and Cynthia Woods Mitchell Institute for 
Fundamental Physics $\&$ Astronomy, Texas A. $\&$ M. University, 
Department of Physics and Astronomy, 4242 TAMU, College Station,
TX 77843, USA, email: yi.yang@weizmann.ac.il}
\altaffiltext{2}{Department of Particle Physics and Astrophysics, 
Weizmann Institute of Science, Rehovot 76100, Israel}
\altaffiltext{3}{Purple Mountain Observatory, Chinese Academy 
of Sciences, Nanjing 210008, China}
\altaffiltext{4}{European Organisation for Astronomical Research 
in the Southern Hemisphere (ESO), Karl-Schwarzschild-Str. 2, 85748 
Garching b.\ M{\"u}nchen, Germany}
\altaffiltext{5}{Space Telescope Science Institute, Baltimore, 
MD 21218, USA}
\altaffiltext{6}{Department of Physics, Florida State University, 
Tallahassee, Florida 32306-4350, USA}
\altaffiltext{7}{Department of Physics and Astronomy, University of 
Sheffield, Hicks Building, Hounsfield Road, Sheffield S3 7RH, UK}
\altaffiltext{8}{Physics Department and Tsinghua Center for 
Astrophysics (THCA), Tsinghua University, Beijing, 100084, China}
\altaffiltext{9}{Department of Astronomy and McDonald Observatory, 
The University of Texas at Austin, Austin, TX 78712, USA}
\altaffiltext{$\dagger$}{Royal Society Research Fellow}

\begin{abstract}
Optical polarimetry is an effective way of probing the environment 
of supernova for dust. 
We acquired linear {\it HST} ACS/WFC polarimetry in bands $F475W$, 
$F606W$, and $F775W$ of the supernova (SN) 2014J in M82 at six 
epochs from $\sim$277 days to $\sim$1181 days after the $B$-band 
maximum. The polarization measured at day 277 shows conspicuous 
deviations from other epochs. These differences can be attributed 
to at least $\sim$ 10$^{-6} M_{\odot}$ of circumstellar dust located at a distance 
of $\sim5\times10^{17}$ cm from the SN. The scattering dust grains 
revealed by these observations seem to be aligned with the dust in 
the interstellar medium that is responsible for the large reddening 
towards the supernova. The presence of this circumstellar dust sets 
strong constraints on the progenitor system that led to the explosion 
of SN\,2014J; however, it cannot discriminate between single- and 
double-degenerate models. 
\end{abstract}

\section{Introduction \label{intro}}

The explosions of type Ia supernovae (SNe) are powered by the thermonuclear 
runaway of ($\sim$1$M_{\odot}$) carbon/oxygen white dwarfs (C/O WDs, 
\citealp{Hoyle_etal_1960}). The homogeneity of type Ia SNe lightcurves (i.e., 
\citealp{Barbon_etal_1973, Elias_etal_1981}), and the correlation between 
the decline rate of the light curve and the luminosity at peak 
\citep{Phillips_etal_1993} enables the usage of type Ia SNe as the most 
accurate distance indicators at redshifts ($z$) out to $\sim$2 
\citep{Riess_etal_1998, Perlmutter_etal_1999, Riess_etal_2016}. 
The exact progenitor systems of type Ia SN explosions remain unknown. 

Some evidence suggests a non-degenerate companion scenario in which a 
compact WD accretes matters from a subgiant or a main sequence star. Examples 
include the time evolution of Na D$_2$ features after the $B-$band maximum light 
of SN\,2006X \citep{Patat_etal_2007}, {an excess of blue light from a normal 
type Ia SN\,2012cg at 15 and 16 days before the $B-$band maximum light 
\citep{Marion_etal_2016}, and the UV flash within $\sim$5 days after the 
explosion of iPTF14atg \citep{Cao_etal_2015}, although iPTF14atg is about 3 
magnitudes subluminous compared to a normal Type Ia SN.} 
Very recently, high-cadence photometric observation of the type Ia SN\,2017cbv 
reveals a blue excess during the first $\sim1-5$ days after the explosion 
\citep{Hosseinzadeh_etal_2017}. Although the blue bump in the light curve can 
be explained by the SN ejecta interacting with a subgiant star, it could also 
be due to interaction with CSM or the presence of nickel in the outer 
ejecta \citep{Hosseinzadeh_etal_2017}. 

Other observations favor a double degenerate scenario featuring the 
merger of two WDs \citep{Iben_Tutukov_1984, Webbink_1984}, see, for example, 
SN\,2011fe \citep{Nugent_etal_2011, Bloom_etal_2012}. Observations also 
excluded any luminous red giant companion (see, for example, \citealp{Li_etal_2011}), 
but the missing companion could also be M dwarfs \citep{Wheeler_etal_2012}. 
For the first few days after the explosion, 
a collision between material ejected by the SN and a non-degenerate companion 
star would produce optical/UV emission in excess of the rising luminosity from 
radioactive decay \citep{Kasen_2010}. In particular, monitoring of three 
photometrically normal type Ia SNe with the Kepler satellite during their entire rising phase 
\citep{Olling_etal_2015} shows no evidence of interaction between SN ejecta and 
circumstellar matter (CSM) or companion stars, thus ruling out the possibility 
of red giants or larger companions predicted by single degenerate models. 
The absence of CSM around type Ia SNe supports double degenerate progenitor models; 
however, searches for CSM around type Ia SNe are difficult, and the results have 
been in most cases inconclusive. 
{Deep {\it HST} imaging of type Ia SN remnant SNR 0509-67.5 in the Large 
Magellanic Cloud found no signs of a surviving ex-companion star. Searches for 
surviving companions of the progenitor have excluded all giant and subgiant 
companions for SN\,1006 \citep{Gonzalez-Hernandez_etal_2012, Kerzendorf_etal_2017_SN1006}, 
and companions with $L\textgreater10 \ L_{\odot}$ for SN\,1604 (Kepler supernova, 
\citealp{Kerzendorf_etal_2014_kepler}). These results strongly disfavoring 
the single-degenerate models \citep{Schaefer_Pagnotta_2012}. However, see a possible 
exception for SN\,1572 \citep{Ruiz-Lapuente_etal_2004}.} 

The merger of two compact stars is a very asymmetric process, which should lead 
to a strong polarimetric signature \citep{Bulla_etal_2016}. By contrast, 
observations consistently find 
a lack of intrinsic polarization before optical maximum \citep{Wang_Wheeler_2008, 
Maund_etal_2013}, which seems to cast doubt on the double degenerate models 
\citep{Wang_Wheeler_2008, Rimoldi_etal_2016}. Quantifying the amount of CSM is of 
high importance for the understanding of the progenitor systems of type Ia SNe. 

Moreover, better estimation of interstellar extinction reduces 
systematic uncertainties. Characterization of dust in the diffuse interstellar 
medium (ISM) relies heavily on the observed wavelength dependence of extinction 
and polarization \citep{Voshchinnikov_etal_2012, Patat_etal_2015}. The observed 
wavelength dependence of interstellar extinction $R_V$ contains information on 
both the size and composition of the grains. The value of $R_V = 3.1$ 
\citep{Cardelli_etal_1989} has often been considered the Galactic standard, but 
with a range from 2.2 to 5.8 \citep{Fitzpatrick_1999} for different lines of 
sight. There is increasing evidence that extinction curves towards type Ia SNe 
systematically favor a steeper law ($R_V \textless3$, see, i.e., 
\citealp{Nobili_etal_2008}, and \citealp{Cikota_etal_2016} for a summary of $R_V$ 
results of earlier studies). This discrepancy has remained unexplained. It is 
very important to understand whether systematically low $R_V$ values towards type 
Ia SNe are caused by systematic differences between the dust compositions of the 
host galaxies. 

{\citet{Wang_2005} and \citet{Patat_etal_2006} have proposed that circumstellar 
dust scattering may be a solution to the surprisingly low $R_V$ values towards 
type Ia SNe, due to a time-dependent scattering process. \citet{Goobar_2008} 
confirmed these results without including the time-dependent radiative transfer 
effect.} The effect on $R_V$ and the 
light curve shape, however, also depends on the large-scale geometrical 
configuration and the properties of the dust grains 
\citep{Amanullah_Goobar_2011, Brown_etal_2015}. For example, recent observations 
of the highly reddened SN\,2014J in M82 have found no convincing evidence of the 
presence of circumstellar dust {(\citealp{Patat_etal_2015, Brown_etal_2015, 
Johansson_etal_2017, Bulla_etal_2018}}, see, however, 
\citealp{Foley_etal_2014, Hoang_etal_2017}). 

Observations in polarized light and its time evolution can be an effective way 
of studying the CSM. Type Ia SNe have low intrinsic polarization in broad-band 
observations ($\lesssim0.2\%$, \citealp{Wang_Wheeler_2008}), whereas the 
scattered light from CSM can be highly polarized. 
{The maximal degree of linear polarization ($p_{\rm max}$) of light 
scattered by dust can reach $\sim$50\% in the $V-$band as reported by, 
e.g., \citet{Sparks_etal_2008} for the light echo from the dusty nebula 
around the eruptive star V838 Mon and by \citet{Kervella_etal_2014} for 
the nebula which contains the $\delta$ Cepheid RS Pup). More typical 
values of $p_{\rm max}$ in the Milky Way are 20\%$-$30\% \citep{Draine_2003b}. 
Theoretical models \citep{Mathis_etal_1989} suggest that interstellar dust 
grains are loose structures with high porosity. This is confirmed by probes 
of cometary dust collected by space and ground-based missions 
(e.g., \citealp{Schulz_etal_2015, Noguchi_etal_2015}), which according 
to \citet{Greenberg_1986} is a proxy of ISM dust. Polarimetry of cometary 
dust found $p_{\rm max}$ values of 10\%-30\% (i.e., see Figure 1 
of \citealp{Petrova_etal_2000} and a review by \citealp{Mann_etal_2006}), 
comparable to the values in the Milky Way ISM. In laboratory experiments 
with analog fluffy aggregates, polarizations in the 50\%$-$100\% range 
were measured \citep{Volten_etal_2007}. In a very recent study, 
\citet{Sen_etal_2017} concluded that, over the range in porosity of 
0\%$-$50\%, $p_{\rm max}$ varies nonmonotonically and can reach or 
exceed 60\%.} 
For a spatially unresolved source, 
the scattered light can contribute significantly to the total integrated 
light and associated distance estimates. In addition, polarization of the 
integrated light can evolve rapidly after maximum light \citep{Wang_Wheeler_1996}. 
{The fraction of polarized flux from any nonaxisymmetric circumstellar dust 
increases substantially as the SN dims and scattered photons (often from light at 
optical maximum) contribute significantly to the SN light curve at late phases. 
The actual situation may be more complicated as the dust distribution can be 
more uniform around the SN than the often assumed single clump, and the 
effect on the polarization and the light curve may be less dramatic. In 
general, the effect is qualitatively stronger in the blue than in the red 
due to the higher scattering opacity in the blue.} 

SN\,2014J was discovered on Jan. 21.805 UT \citep{Fossey_etal_2014,
Ma_etal_2014}, and the first light has been constrained to be Jan. 14.75 
UT \citep{Zheng_etal_2014,Goobar_etal_2015}. SN\,2014J reached its $B$-band 
maximum on Feb. 2.0 UT (JD 2,456,690.5) at a magnitude of 11.85$\pm$0.02 
\citep{Foley_etal_2014}. Exploding in the nearby starburst galaxy M82 at 
a distance of 3.53$\pm$0.04 Mpc \citep{Dalcanton_etal_2009}, SN\,2014J was 
the nearest SN since SN\,1987A. The relative proximity of SN\,2014J allows 
continuous photometric and spectroscopic observations through late phases 
\citep{Lundqvist_etal_2015, Bonanos_etal_2016, Porter_etal_2016, 
Sand_etal_2016, Srivastav_etal_2016, Johansson_etal_2017, Yang_etal_2017b}. 
SN\,2014J suffers from heavy extinction and is located behind a large amount 
of interstellar dust \citep{Amanullah_etal_2014}. There is ample 
evidence that the strong extinction is caused primarily by interstellar 
dust {\citep{Patat_etal_2015,Brown_etal_2015, Bulla_etal_2018}}; 
however, high resolution 
spectroscopy does show strong evidence of time evolving K{\sc I} lines that 
can be understood as due to photo-ionization of material located at a 
distance of about 10$^{19}$ cm from the SN \citep{Graham_etal_2015}. 
Moreover, numerous Na, Ca and K features along the SN-Earth line of sight 
were detected \citep{Patat_etal_2015}. No positive detection of any 
material at distances within 10$^{19}$ cm has been reported for SN\,2014J, 
but see \citet{Foley_etal_2014, Brown_etal_2015, Bulla_etal_2016} for an 
alternate view. In this paper, 
we present our late-time {\it HST} imaging polarimetry of SN\,2014J and 
derive from it the amount of circumstellar dust around SN\,2014J. 

\section{Observations and Data Reduction}
The {\it HST} WFC/ACS camera has a polarimetry mode which allows for accurate 
imaging polarimetry. The filter-polarizer combinations selected by us have 
recently been calibrated \citep{Avila_etal_2017}. We used the Advanced Camera 
for Surveys/Wide Field Channel (ACS/WFC) on board the {\it HST} to observe 
SN\,2014J in imaging polarization mode at six epochs (V1-V6) under multiple 
{\it HST} programs: GO-13717 (PI: Wang), GO-14139 (PI: Wang), and GO-14663 
(PI: Wang). The observations were taken with three different filters: $F475W$ 
(SDSS $g$), $F606W$ (broad $V$), and $F775W$ (SDSS $i$), each combined with 
one of the three polarizing filters: POL0V, POL60V, and POL120V oriented at 
relative position angles of $0^{\circ}$, $60^{\circ}$, and $120^{\circ}$, 
respectively. A log of observations is presented in Table~\ref{Table_log}. 
Multiple dithered exposures were taken at each observing 
configuration to allow for drizzling of the images. Exposure times ranged 
from 30 s with $F775W$ on day 276 to 1040 s with $F475W$ on day 1181. 

The {\it HST} data were reduced following the usual routine of drizzling 
to remove artifacts and cosmic rays. For 
each bandpass and polarizer, one combined image was prepared. Bright HII 
regions in the field-of-view (FOV) were used to align exposures in different 
bandpass$+$polarizer combinations and epochs through {\it Tweakreg} in the 
{\it Astrodrizzle} package \citep{Gonzaga_etal_2012}. 
{The polarizers contain a weak optical lens which corrects the optical 
focus for the presence of bandpass$+$polarizer filters in the light path. 
Large scale distortions introduced by this weak optical lens have been 
removed using the {\it Astrodrizzle} software.} 
All images were aligned to better than 0.25 pixels in both $x$ and $y$ directions. 
This is compatible with the small scale distortion ($\pm$0.3 pixel) in 
the images caused by slight ripples in the polarizing material (see the 
ACS Data Handbook, \citealp{Lucas_etal_2016}). 

\begin{figure}[!htb]
\epsscale{1.0}
\plotone{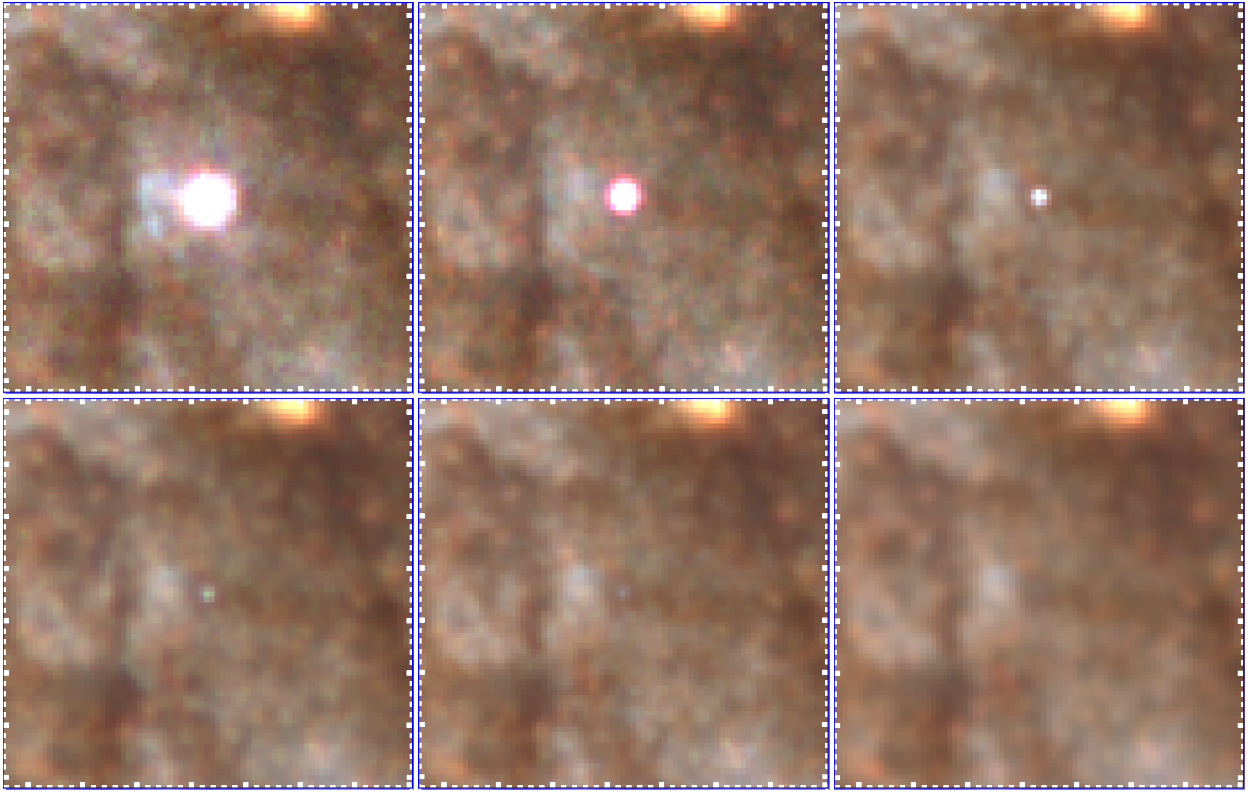}
\caption{\footnotesize Color images of SN\,2014J from {\it HST} ACS/WFC $F475W$, $F606W$, 
and $F775W$ observations on days 277 (upper left), 416 (upper middle), 649 
(upper right), 796 (lower left), 985 (lower middle), and 1181 (lower right) 
after maximum light. North is up, east is left, and the distance between 
big tick marks corresponds to 0$\arcsec$.5 or 8.6 pc projected on the 
plane of the sky. Reflection of SN light by the dust between the SN and 
the observer creates arcs of light echoes which are propagating with time. 
There may also be unresolved light echoes at distances so close to the 
central SN that even the {\it HST} cannot resolve them, but imaging polarimetry 
can still detect their presence. 
\label{Fig_image}
} 
\end{figure}

The absolute throughput values of bandpass$+$polarizer combinations 
listed in the {\it Synphot \footnote[1]{
http://www.stsci.edu/institute/software\_hardware/stsdas/synphot}
} software does not match those found in on-orbit calibrations. 
Correction factors by \citet{Cracraft_Sparks_2007} based on 
on-orbit calibration programs were used to remove the instrumental 
polarization. The scaling factors ($C_{POL*V}$) have been applied 
to images obtained with each polarizer: 
$r(POL*V) = C_{POL*V} * Im(obs)_{POL*V}$. The remaining instrumental 
polarization can still be as much as $\sim$1\%, 
and the instrumental polarization has been observed to vary with roll angle 
(i.e., see \citealp{Cracraft_Sparks_2007} and 
\citealp{Lucas_etal_2016}). To improve the measurement precision, we 
use bright sources in the field (for visits V1 and V2) to 
monitor the stability of the instrumental polarization. The roll 
angles in the subsequent observing epochs were set to be equal to 
or 180$^{\circ}$ different from the roll angles in V1 and V2. We 
discuss this further in Section 3. 

\subsection{Measuring the degree of polarization}
We deduced the Stokes ($I$,$Q$,$U$) from the observations as follows: 
\begin{equation}
\begin{aligned}
I = \frac{2}{3} [r(POL0) + r(POL60) + r(POL120)], \\
Q = \frac{2}{3} [2r(POL0) - r(POL60) - r(POL120)], \\ 
U = \frac{2}{\sqrt{3}} [r(POL60) - r(POL120)], 
\end{aligned}
\label{Eqn_stokes}
\end{equation}
where $I$, $Q$, and $U$ are standard notation of the components of the 
Stokes vector. 
Flux measurements were made with a circular aperture of 0.$\arcsec$15 
(3 pixels in the ACS/WFC FOV) to reduce the contamination from the 
extremely non-uniform background. Aperture corrections were calculated 
with the ACS/WFC encircled energy profile for each bandpass according 
to \citet{Sirianni_etal_2005}. We perform the measurements of the SN 
on the images obtained by each polarizer $r(POL*V)$. We also deduce the 
Stokes $I$, $Q$, $U$ maps using Equation \ref{Eqn_stokes}, integrating 
within the aperture centered at the SN on the Stokes $I$, $Q$, 
$U$ maps. In both cases, the background has been estimated by choosing 
the same inner and outer radii as used by \citet{Yang_etal_2017b}. The 
two approaches agree within the uncertainties when the signal-to-noise 
($S/N$) ratio on each $r(POL*V)$ is S/N $\textgreater$50. 
Figure~\ref{Fig_image} presents a color composite image of SN\,2014J 
consisting of the Stokes $I$ data for each bandpass and epoch. The 
images show resolved light echoes expanding over time, which were 
first identified by \citet{Crotts_2015}. We only remark here that these 
multiple light echoes are produced by dust clouds at a distance about 
100 pc to 500 pc away from SN\,2014J. The dust in those sheets is unlikely to be related 
to the evolution of the SN progenitor. Detailed studies of these resolved 
light echoes were performed in the same {\it HST} data as those used for 
the present study and can be found in \citet{Yang_etal_2017a}. 

The degree of polarization and the polarization position 
angle can be derived as: 
\begin{equation}
p\%=\frac{\sqrt{Q^2+U^2}}{I} \times \frac{T_{par}+T_{perp}}{T_{par}-T_{perp}} \times 100 \%
\label{Eqn_p}
\end{equation}
\begin{equation}
PA=\frac{1}{2}\mathrm{tan}^{-1}\bigg( \frac{U}{Q} \bigg)+PA\_V3+\chi
\label{Eqn_pa}
\end{equation}
The SN fluxes measured in the different `bandpass$+$polarizer' combinations 
were then converted to polarization measurements following the {\it HST} ACS 
manual \citep{Avila_etal_2017} and earlier work \citep{Sparks_Axon_1999}. 
The cross-polarization leakage is insignificant for visual polarizers 
\citep{Biretta_etal_2004}. The factor containing the parallel and perpendicular 
transmission coefficients $(T_{par}+T_{perp})/(T_{par}-T_{perp})$ is about unity 
and has been corrected in our data reduction. The degree of polarization ($p\%$) 
is calculated using the Stokes vectors. These corrections together with the 
calibration of the source count rates vectorially remove the instrumental 
polarization of the WFC ($\sim 1\%$). The polarization position angle ($PA$) 
is calculated using the Stokes vectors and the roll angle of the {\it HST} 
spacecraft ($PA\_V3$ in the data headers) as shown in Equation \ref{Eqn_pa}. 
Another parameter, called $\chi$, containing information about the camera 
geometry which is derived from the design specification, has been considered 
when solving the matrix to deduce the Stokes vectors. For the WFC, 
$\chi=-38.2^\circ$ \citep{Lucas_etal_2016}. 

\subsection{Errors in Polarimetry}
The classical method proposed by \citet{Serkowski_1958, Serkowski_1962} is often 
used for the determination of the polarization and associated uncertainties. 
\citet{Montier_etal_2015} investigated the statistical behavior of basic 
polarization fraction and angle measurements. 
{We use Equations~\ref{Eqn_pe} and \ref{Eqn_pae} to describe the uncertainty 
of $p$ and $PA$, where $\sigma_I, \sigma_Q, \sigma_U$ denotes the associated errors in 
individual measurement of the Stokes $I,Q,U$; $\sigma_{QU}, \sigma_{IQ}, \sigma_{IU}$ 
denotes the covariance between the associate Stokes parameters.} 
The detailed derivation is available in Appendix F of 
\citet{Montier_etal_2015}. 
\begin{equation}
\sigma_p ^2 \ = \  \frac{1}{p^2 I^4} \times \big{(} Q^2 \sigma_Q ^2 + U^2 \sigma_U ^2 + p^4 I^2 \sigma_I ^2 
+ 2QU \sigma_{QU} - 2IQp^2 \sigma_{IQ} -2IUp^2 \sigma_{IU}^2 \big{)}
\label{Eqn_pe}
\end{equation}
\begin{equation}
\sigma_{P.A} \ = \  \sqrt{\frac{Q^2\sigma_U ^2 + U^2\sigma_Q ^2 -2QU\sigma_{QU}}{Q^2\sigma_Q ^2 + U^2\sigma_U ^2 +2QU\sigma_{QU}}} 
\times \frac{\sigma_p}{2p} \  rad
\label{Eqn_pae}
\end{equation}
The Stokes $I$ component gives the total intensity of the source. The AB 
magnitudes of the SN were obtained by applying the ACS/WFC zeropoints. 

The degree of polarization and the magnitudes of the SN in different filter 
bands are shown in Table~\ref{Table_pol}. The other sources of data used in 
this paper include three epochs of observations \citep{Patat_etal_2015} using 
the polarimetric mode of the Calar Alto Faint Object Spectrograph (CAFOS, 
see \citealp{Patat_etal_2011}) instrument at the 2.2 m telescope in Calar Alto, 
Spain. The spectropolarimetry used the low-resolution B200 grism coupled with a 
1.5$\arcsec$ slit, giving a spectral range 3300-8900 \AA, a dispersion 
of $\sim$ 4.7 \AA/pix, and a full width half maximum (FWHM) resolution of 
21.0 \AA. Spectropolarimetry from Calar Alto was obtained on Jan 28 (day -6), 
Feb 03 (day 0, already published in \citealp{Patat_etal_2015}) and Mar 08 (day 33) 2014. 
We also used broad-band polarimetry taken with the Hiroshima One-shot Wide-field 
Polarimeter (HOWPol, \citealp{Kawabata_etal_2008}) around optical maximum as 
published by \citet{Kawabata_etal_2014}. 

\section{Analysis}
\begin{figure}[!htb]
\epsscale{0.65}
\plotone{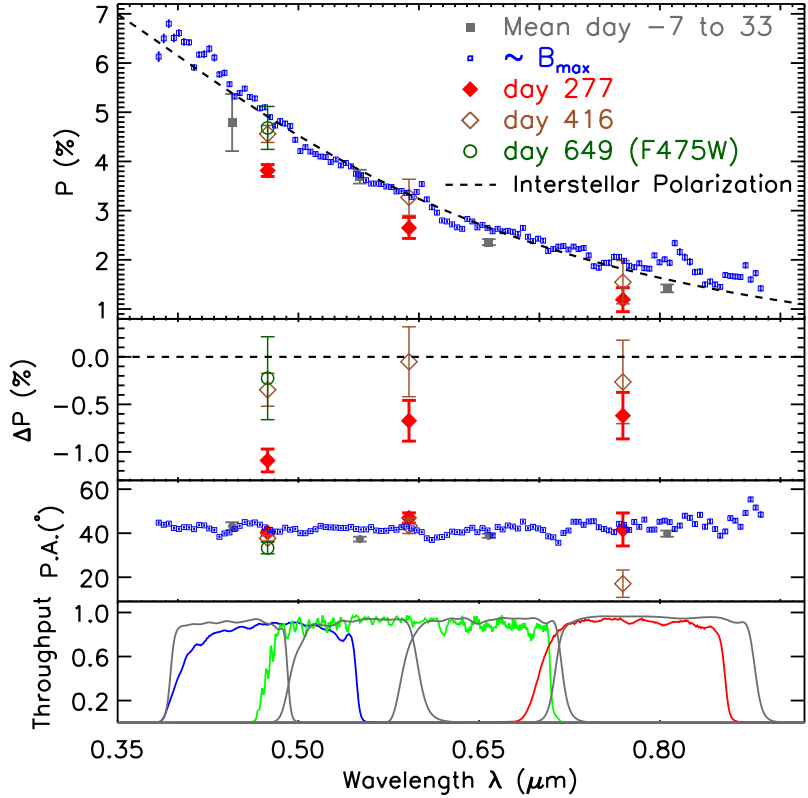}
\caption{\footnotesize From top to bottom: the first panel presents the optical imaging 
polarimetry of SN\,2014J taken with {\it HST} ACS/WFC on day 277, day 416, 
and day 649, compared with earlier broad-band polarization between  day -7 and day 33 (gray, solid 
squares, \citealp{Kawabata_etal_2014}) and spectropolarimetry near $B-$band maximum (blue, open 
squares, \citealp{Patat_etal_2015}). The dashed line presents the `Serkowski 
law' fit of the interstellar polarization; the second panel gives the 
difference between our {\it HST} polarimetry and the interstellar 
polarization; the third panel displays the corresponding polarization 
position angles; the bottom panel illustrates the filter transmission curves 
for the broad-band polarimetry \citep{Kawabata_etal_2014} (gray lines), 
and the {\it HST} $F475W$ (blue line), broad $F606W$ (green line), and 
$F775W$ (red line) filter band measurements. 
The {\it HST} data on day 277 exhibit 
a conspicuously different degree of polarization in all three filter bands 
compared to the other data sets. At later epochs, the polarization returns 
to the values at maximum light. 
\label{Fig_pol_lambda}
}
\end{figure}

Figure~\ref{Fig_pol_lambda} presents the wavelength dependence and time 
evolution of the new {\it HST} data points together with ground-based 
polarimetry. The {\it HST} data can be compared to ground-based 
polarimetry acquired around optical maximum to study the temporal 
evolution of the polarization. Broad-band polarimetric observations of 
SN\,2014J taken on Jan 22.4 (-11 days relative to $B$-band maximum), 
Jan 27.7 (-6 days), Feb 16.5 (+14 days), Feb. 25.6 (+23 days) and 
Mar 7.8 (+33 days) detected no variability \citep{Kawabata_etal_2014}. 
Spectropolarimetry on Jan 28 (-6 days), Feb 03 (+0 day), and Mar 08, 2014 
(+33 days) indicates no temporal evolution either \citep{Patat_etal_2015}. 
The continuum polarization of SN\,2014J reaches about 6.6\% at 0.4 $\mu m$, 
and the variability in ground-based data was less than 0.2\%, except at 
the bluest end where the data were noisy but are still consistent with 
constancy \citep{Patat_etal_2015}. At the 0.2\% level, the intrinsic 
polarization of the SN becomes significant \citep{Wang_Wheeler_2008}. 
This makes it difficult to determine the contribution from circumstellar 
dust. We conclude that the overall high level of polarization at early 
times is due to interstellar dust, and that there is no detectable 
variability at early times down to the 0.2\% level. 

\subsection{Interstellar polarization}
The ``Serkowski Law'' provides an empirical wavelength 
dependence of optical/near infrared (NIR) interstellar polarization 
\citep{Serkowski_etal_1975}. It can be written as: 
\begin{equation}
p(\lambda)/p_{\mathrm{max}} = \mathrm{exp}[-K\ \mathrm{ln}^2 (\lambda_{\mathrm{max}} / \lambda)], 
\label{Eqn_Ser}
\end{equation}
where $\lambda_{\mathrm{max}}$ is the wavelength of the maximum polarization 
$p(\lambda_{\mathrm{max}})$ and $K$ is a parameter describing the width of the 
polarization peak. 
We fitted this relation to optical spectropolarimetry at 
maximum light. 
The interstellar 
polarization wavelength dependence towards SN\,2014J exhibits a very steep 
increase from the red to the blue \citep{Kawabata_etal_2014, Patat_etal_2015}. 
The position of the polarization peak cannot be determined due to the lack of 
UV data. Therefore, we employ the canonical value $K=1.15$ according to 
\citet{Serkowski_etal_1975} and obtain a reasonable fit with 
$\lambda_{\mathrm{max}} = $0.25 $\mu m$ and $p(\lambda_{\mathrm{max}})=$8.1\%. 
{Our fitting to Serkowski's law, together with the polarimetry of SN\,2014J were 
shown in the first panel of Figure~\ref{Fig_pol_lambda}}. 
Extrapolation to the effective wavelengths of the $F475W$, $F606W$ and $F775W$ filters 
yields values of 4.9\%, 3.3\% and 1.8\% respectively for the interstellar polarization. 

\subsection{Polarimetry of light scattered from a SN \label{Section_model}}
{In the {\it HST} data from day 277, the $F475W$-band degree of polarization 
has changed from 4.9\% near maximum light to 3.8\%, and no obvious change in 
$PA$ has been observed.} A stability check of the {\it HST} polarimetry will 
be presented in Section 3.3. The $F475W$-band data have 
the highest $S/N$. The data in the $F606W$ and $F775W$ bands also show 
different degrees of polarization. The data on day 416, however, are consistent 
with those from maximum light. Polarimetry at later epochs suffers from larger 
uncertainties as the SN fades; however, it is still broadly consistent with the 
interstellar polarization. \citet{Sparks_Axon_1999} fitted the errors of the 
polarization degree and the polarization position angle with the average $S/N$ 
ratio and the degree of polarization: 
\begin{equation}
\begin{aligned}
\mathrm{log}_{10} (\sigma_{p} / p) = - 0.102 - 0.9898 \mathrm{log}_{10} (p \langle \mathrm{S/N} \rangle_{i}) \\
\mathrm{log}_{10} \sigma_{PA} = 1.415 - 1.068 \mathrm{log}_{10} (p \langle \mathrm{S/N} \rangle_{i})
\end{aligned}
\label{Eqn_err}
\end{equation}
For example, exposures at each polarizer achieving 
$\langle \mathrm{S/N} \rangle_{i} \sim 500$ yield relative uncertainties 
$\sigma_{p}/p$ = 3.3\%, 4.9\%, and 9.0\% in the $F475W$, $F606W$, and $F775W$ 
bandpasses, respectively. For $\langle \mathrm{S/N} \rangle_{i} \sim 100$, the 
corresponding values are $\sigma_{p}/p$ = 16\%, 24\%, and 44\% in the $F475W$, 
$F606W$, and $F775W$ bandpasses, respectively. 
{The exposure time in the $F475W$ band at later epochs was longer, and 
the average $S/N$ ratio for the SN point source is estimated as 700, 450, 190, 
100, 40, and 30, leading to values in $(\sigma_{p} / p)$ of 3.1\%, 4.6\%, 11\%, 
21\%, 50\%, and 70\%, from V1 to V6, respectively. The fractional errors are 
also in good agreement with the errors derived with Equation~\ref{Eqn_pe} and 
presented in Table~\ref{Table_pol}.} 
The polarization position angles ($PA$) at 
all visits are broadly consistent with the average polarization position angle 
42.2$\pm$0.3 deg derived around maximum light \citep{Patat_etal_2015}. 

Differences in observed polarization on day 277 can be explained with a 
non-uniform distribution of circumstellar dust in the vicinity of SN\,2014J. 
Modeling the observed polarization in terms of dust scattering of SN light is usually an 
ill-defined problem due to the lack of knowledge about the geometric 
distribution of the dust and its absorption and scattering properties. A 
unique solution is usually very difficult to achieve; however, important 
constraints can be deduced based on simple and robust models. 

The most efficient configuration for producing polarized light is given by a 
single dust clump near the location of the SN but offset from the SN on or 
close to the plane of the sky. In such a configuration, the light incident 
on the dust clump is scattered near 90$^{\circ}$ and can be polarized at the 
50-100\% level. The degree of polarization depends on the details of the 
geometry and optical depth of the dust clump. For simplicity, and without 
loss of much generality, the amount of scattered light can be written as the 
following equation: 
\begin{equation}
L_{scat}(t)\ = \ \tau \frac{\delta \Omega}{4\pi}\Phi(\theta) \int{L(t-t_e) K(t-t_{d})}dt_e,
\label{Eqn_scatter1}
\end{equation}
where $t$ and $t_e$ give the time of observation and the time since 
SN explosion, respectively, {$\tau$ is the optical depth along the 
scattering direction in the circumstellar cloud}, 
$\delta \Omega$ is the solid angle the clump subtends toward the SN, $L(t)$ is 
the luminosity of the SN as a function of time, $t_d$ denotes the light travel time 
from the SN to the center of the dust clump, $\theta$ gives the scattering angle, 
and $\Phi(\theta)$ is the scattering phase function. We assume that dust 
scattering follows the Henyey--Greenstein phase function \citep{Henyey_Greenstein_1941}: 
\begin{equation}
\Phi(\theta) =   
\frac{1}{4\pi} \frac{1-g^2}{(1+g^2-2gcos\theta)^{3/2}}, 
\label{Eqn_phasefunction}
\end{equation}
where $g=\overline{cos\theta}$ is a measure of the degree of forward
scattering and computed by \citet{Laor_Draine_1993}. 
The function $K$ is determined by the details of the dust distribution. It 
reduces to an infinitely narrow Dirac $\delta$-function for an infinitely thin 
layer of dust lying on the surface of the light travel iso-delay surface (see 
\citealp{Patat_2005}). For a more realistic distribution, $K$ reduces to a 
broader function whose width characterizes the radial extent of the clump. 
The lack of a precise geometric model of the dust clump leads us to approximately 
describe the scattering properties of the clump with a) an infinitely narrow Dirac 
$\delta$-function, and b) a Gaussian function of the form 
$K(t)\ = \ \frac{1}{\sqrt{2\pi}\sigma_t} \exp(-\frac{t^2}{2\sigma_t^2})$. 
Here $\sigma_t\times c$ characterizes the radial extent of the clump, and 
$\tau$ can be the average optical depth of the clump which is linearly related 
to the average column depth in the case of an optically thin clump. {In the 
following, we use the more restrictive Dirac $\delta$-function assumption to deduce 
the minimal amount of dust responsible for the late-time variations in polarization. 
In addition, we calculate this quantity also for a radially extended dust clump 
approximated by a Gaussian function with $\sigma = 20$ light days.}  

The degree of polarization is then 
\begin{equation}
p\ = \frac{L_{scat}(t)}{L(t)+L_{scat}(t)} {\Theta(\theta)},
\label{Eqn_scatter2}
\end{equation}
where $\Theta(\theta)$ is the polarization of light scattered with scattering angle 
$\theta$. We adopt the Mie scattering \citep{Mie_1976} model for dust particles of 
radius $a$=0.1 $\mu m$, comparable to the wavelengths of the filter bands. 
The scattering phase functions and optical properties of dust particles were 
calculated using the OMLC Mie Scattering Calculator \footnote[2]{
http://omlc.org/calc/mie\_calc.html}. 

Dust located on the iso-delay light surface for a given epoch will produce 
scattered flux. The total mass of the dust responsible for the scattering gives: 
\begin{equation}
M_{dust}=n_{gr}V_{gr}\rho_{gr}dV,
\label{Eqn_mdust1}
\end{equation}
where $n_{gr}$ is the dust grain number density. $V_{gr}$ describes the volume of a 
single dust grain and can be written as: $V_{gr}=A_{gr} l_{gr}$, {where $l_{gr}$ 
represents the effective length perpendicular to a grain's geometric cross-section with 
an area of $A_{gr}$.} The volume of the dust cloud gives $dV=r^2\mathrm{sin}\theta d\theta d\phi dr$ where 
$r=c{t_d}/(1-\mathrm{cos}\theta)$ gives the distance from the SN to a dust cloud, 
and $t_d$ denotes the time within which the SN radiation reaches the dust cloud. 
The optical depth of this dust cloud can be expressed as follows: 
\begin{equation}
\tau = n_{gr} A_{gr} Q_{ext} dr,  
\label{Eqn_tau1}
\end{equation}
{where $Q_{ext}$ gives the extinction efficiency for dust grains.} 
Under the assumption of an infinitely narrow Dirac $\delta$-function dust cloud, 
the amount of scattered photons can be expressed as: 
\begin{equation}
L_{scat}(t) = \omega \tau L(t-t_e) \rm{sin} \theta \delta \theta \delta \phi \Phi(\theta). 
\label{Eqn_scat1}
\end{equation}

The amount of polarized scattered light can therefore be used to infer the 
optical depth and mass of the scattering dust cloud. When the light from the 
SN is still dominant over the scattered light by the circumstellar dust cloud, 
i.e., $L(t) \gg L_{scat}(t)$, and recalling Equation~\ref{Eqn_scatter2}, 
{Equation~\ref{Eqn_scat1} can be rewritten as:} 
\begin{equation}
\tau = p \frac{1}{\omega} \frac{L(t)}{L(t-t_e)} \frac{1}{\delta \theta} \frac{1}{\delta \phi} 
\frac{1}{\Phi(\theta)} \frac{1}{\Theta(\theta)} \frac{1}{\rm{sin} \theta}, 
\label{Eqn_tau2}
\end{equation}
{where $\omega$ denotes the grain albedo.} 

The mass of the dust cloud is then given by: 
\begin{equation}
M_{dust}= \tau l_{gr} \rho_{gr} \frac{1}{Q_{ext}} r^2 \mathrm{sin}\theta d\theta d\phi. 
\label{Eqn_mdust2}
\end{equation}
Without knowing the exact shape of the dust grains, it is reasonable to 
replace $l_{gr}$ with the radius of the dust grain, $a$. 
{The grain albedo $\omega$ can be expressed as $\omega = Q_{scat}/Q_{ext}$, 
where $Q_{scat}$ and $Q_{ext}$ give the scattering and the extinction 
efficiency, respectively. We rewrite $\omega Q_{ext}$ as $Q_{scat}$ and 
adopt the values computed for various dust models (see the following paragraph).} 
The lack of knowledge of the geometric size of the dust cloud makes it reasonable 
to assume that the scattering kernel is a function of the geometric width 
of the clump. 
{For a single clump and a thin, Dirac $\delta$-function kernel, 
combining Equation~\ref{Eqn_tau2} and \ref{Eqn_mdust2}, 
we found the following constraints on the dust mass:} 
\begin{equation}
M_{\mathrm{dust}}^{\mathrm{thin}} \geq 1.4 \times 10^{-7} M_\odot \frac{p}{1\%}\left[ \frac{{L(0)}/{L(t_d)}}{1.0\times10^{-4}}\right] 
\left[\frac{c t_{d}/(1-\mathrm{cos}\theta)}{1 \ l.y.}\right]^2 
\frac{1}{Q_{scat}} \ 
\frac{\rho_{gr}}{2.5 g/cm^3}  \ 
\frac{a}{0.1\ \mu m} \  \frac{1}{\Phi(\theta)} \  \frac{1}{\Theta(\theta)}, 
\label{Eqn_scatter3}
\end{equation}
where $p$ is the observed amount of polarization evolution and 
$\rho_{gr}$ is the physical density of the dust grains. 

\begin{figure}[!htb]
\epsscale{0.8}
\plotone{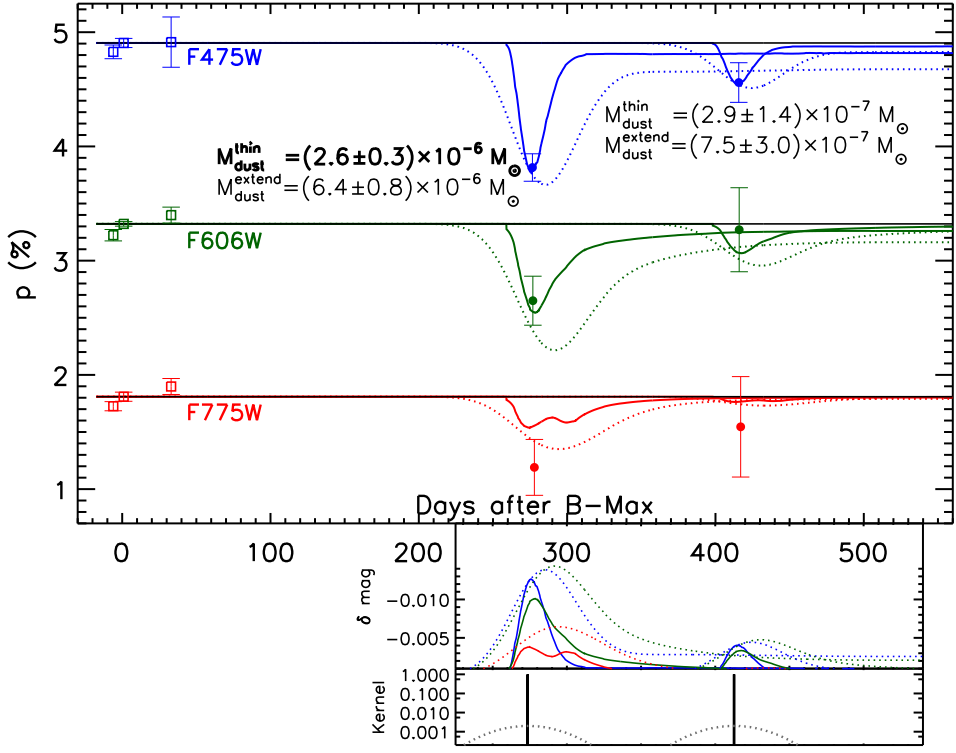}
\caption{\footnotesize Single-dust-clump models of the late-time polarimetry of SN\,2014J. 
In the upper panel, the blue, green, 
and red lines are for $F475W$, $F606W$, and $F775W$-band data, respectively. 
The straight horizontal lines in each color indicate the average 
polarization \citep{Patat_etal_2015} in each filters. The solid lines 
represent the models for an infinitely thin dust distribution, and the dashed 
lines illustrate the models for a radially extended dust clump approximated by 
a Gaussian function with $\sigma=20$ light days. The upper of the two smaller 
panels at the bottom shows 
the expected contribution to the integrated light curves by the hypothetical 
silicate dust clump which can account for the observed polarization evolution. 
The lower of these two panels describes the infinitely thin (Dirac $\delta$-function) and 
the Gaussian dust kernels. All panels share the same time axis. 
\label{Fig_pol_time}
}
\end{figure}

\begin{figure}[!htb]
\epsscale{0.6}
\plotone{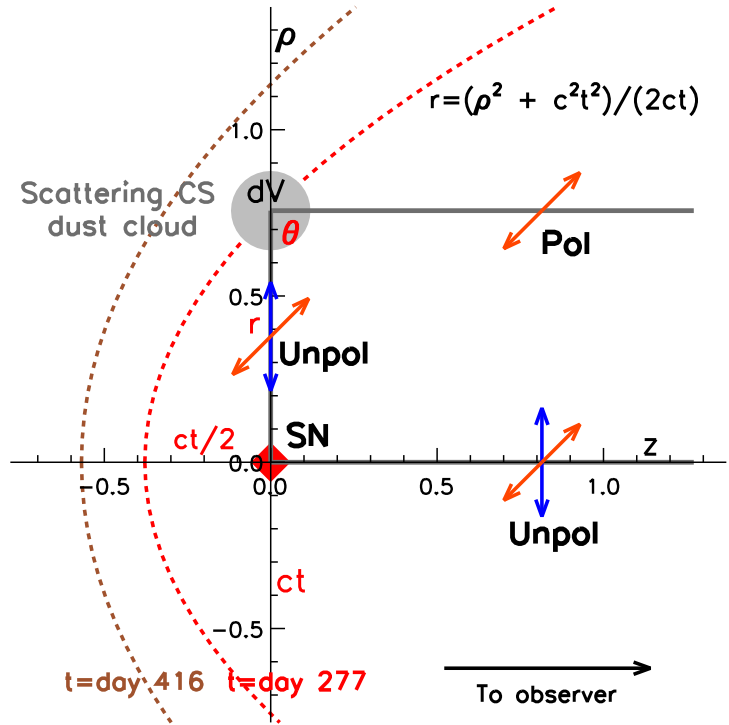}
\caption{\footnotesize Schematic diagram illustrating the geometrical configuration 
of a circumstellar light echo around a supernova. The diagram describes 
the contribution from photons scattered by a circumstellar dust cloud 
at large angle (i.e., $\theta=90^{\circ}$) and the time-variant 
polarization of the SN\,2014J. The abscissa and ordinate represent the 
foreground distance ($z$) in and projected distance on the sky ($\rho$), 
respectively. Both $z$ and $\rho$ are in light years. Paraboloids 
represent the iso-delay light surfaces at different epochs (as labeled), 
`Pol' and `Unpol' denote `polarized light' and `unpolarized light', 
respectively, as seen by the observer located outside the right edge 
of the figure. 
\label{Fig_schem}
}
\end{figure}

{For SN\,2014J, we have identified a strong polarization anomaly at day 277 
after $B-$maximum which shows a polarization that differs in all filter bands 
from the polarization observed around optical maximum and the polarization at 
later times taken by the same program.} 
We applied the above model to the observed data to deduce the amount of 
dust needed to produce the observed polarization at day 277. The results 
for Mie scattering by `astronomical silicate' \citep{Draine_Lee_1984, 
Laor_Draine_1993, Weingartner_Draine_2001} are shown in 
Figure~\ref{Fig_pol_time} for all the three bands. Based on our 
measurement through $F475W$ with the highest $S/N$ ratio, a minimum mass 
of silicate dust of $2.4\times10^{-6} M_{\odot}$ is needed to reproduce 
the observed polarization evolution, at a scattering angle of $114_{-5}^{+5}$ degrees 
with respect to the line of sight. We also considered graphite and Milky 
Way dust, which yield minimal dust masses of 
$(3.6 \pm 0.4) \times10^{-6} M_{\odot}$ and 
$(3.2 \pm 0.4) \times10^{-5} M_{\odot}$, respectively. 
Table~\ref{Table_mass} summarizes the amount of dust inferred from the 
difference in the polarization degree between days 277 and 416. 
The required minimal dust masses were derived from Equation~\ref{Eqn_scatter3}. 
The scattering angles were, then, obtained from the same equation when 
the masses acquire its minimum value; they are sightly dependent on the 
adopted distribution model but always near 90 degrees. 
Uncertainties were estimated through a Monte-Carlo procedure by adding 
Gaussian errors to the parameter values. Because of presence also of 
systematic errors, the resulting error margins are only lower limits of 
the real uncertainties of the single-dust-clump model. 
Figure~\ref{Fig_schem} provides a schematic view of the single dust 
clump model which explains the time-dependent polarization of SN\,2014J. 

A single dust clump close to the plane of the SN leads to the largest 
possible polarization. Any more complex geometric distribution of the 
dust will be less efficient in polarizing scattered light from the SN 
and therefore more dust will be needed to achieve the same degree of 
polarization. Nonetheless, the single dust clump model can provide 
useful insights even for a more complicated geometry such as a 
non-uniform dust distribution. In such a case, the polarization will 
be related to the fluctuations of the column depth of dust to the SN. 

\begin{figure}[!htb]
\epsscale{0.8}
\plotone{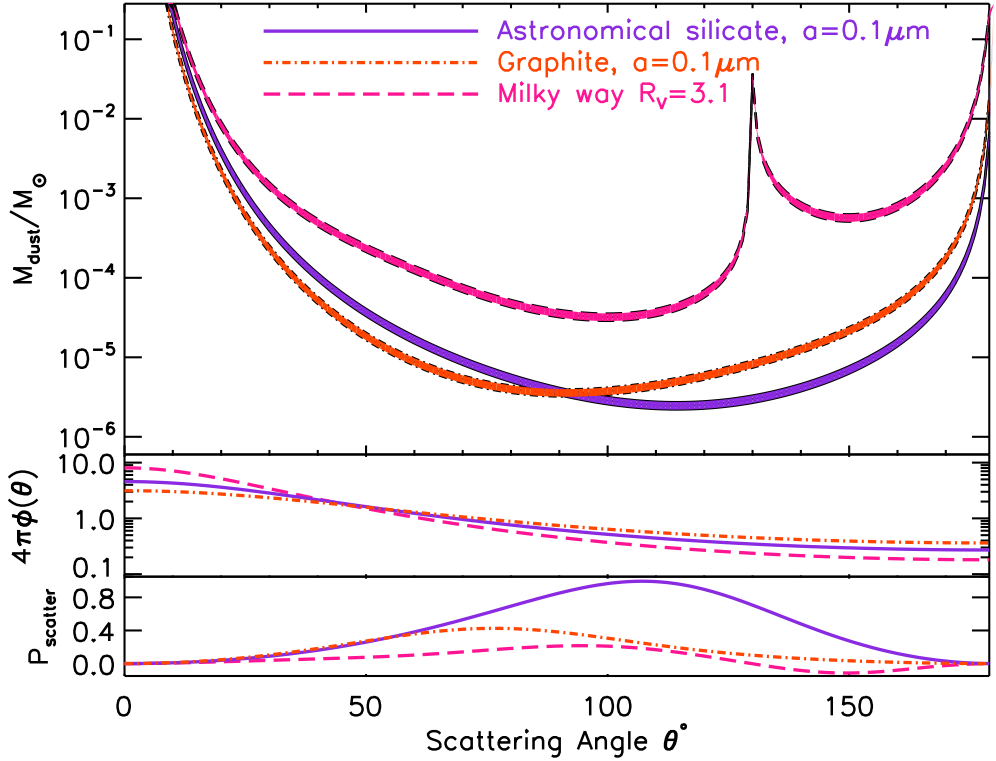}
\caption{\footnotesize The dust mass required to produce the observed level of 
polarization as a function of scattering angle caused by the dust 
clump's position along the iso-delay surface at day 277 (when the 
deviant polarization was measured) and as depicted in Figure~\ref{Fig_schem}, 
which shows the case of the $\theta=90^{\circ}$. 
In the upper panel, the cases 
of silicate, graphite, and Milky Way dust are represented by a 
solid purple, dotted-dashed orange, and dashed pink line, 
respectively. The scattering-angle dependency of scattering phase 
functions and polarization efficiencies obtained from 
\citet{Weingartner_Draine_2001} are overplotted in the middle and 
bottom panel, respectively. 
\label{Fig_pol_angle}
}
\end{figure}

For dust distributed in a torus viewed edge-on, the amount of dust 
needed is $\sim 2\pi/\delta\theta$ times larger than demanded by 
the single dust clump model with an angular size $\delta\theta$. 
Figure~\ref{Fig_pol_angle} presents the amount of dust required to 
account for the observed change in polarization at different 
scattering angles. This allows the single dust clump to move along 
the iso-delay light surface in Figure~\ref{Fig_schem} and provides 
a more universal description of the implied dust mass. 
The minimum amount of dust that is compatible with a torus geometry 
is still consistent with constraints from NIR observations, i.e., 
$10^{-5}M_{\odot}$ inside a radius $1.0\times 10^{17}$ cm 
\citep{Johansson_etal_2017}. If we model the polarization in terms 
of a non-uniform spherical shell, 
the required mass will be larger than or of the order of 
4$\pi/\delta\theta^2$ times that of a single dust clump. 

\subsection{Stability check of the {\it HST} polarimetry} 
{\it HST} has {obtained only few other polarimetric observations 
of point sources that could be used to assess the quality of the 
observations of SN\,2014J. Therefore, in order to test the stability 
of {\it HST} polarimetry, we have also measured the polarization of a 
number of stars and nebular sources in the surrounding {\it HST} WFC 
field. {We assume that the polarization of the field sources other 
than SN\,2014J is due to polarization from foreground dust, and, 
therefore, time invariant.} These stars and nebulae are identified in 
Figure~\ref{Fig_image_blob}. The evolution of their polarization 
between days 277 and 416 is visualized in Figure~\ref{Fig_qu_blob}. 

\begin{figure}[!htb]
\epsscale{0.8}
\plotone{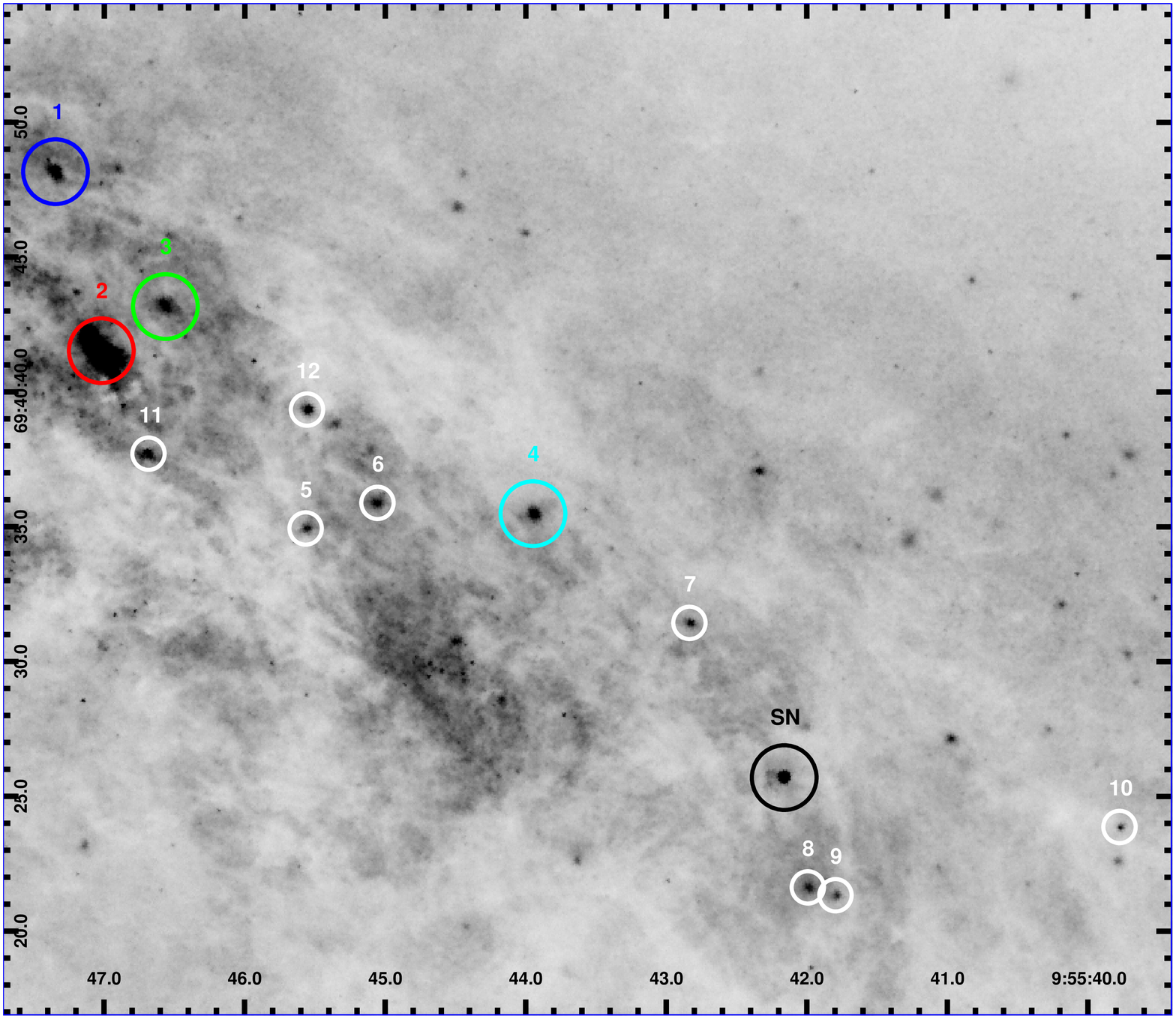}
\caption{\footnotesize The bright sources in the {\it HST} images used for 
determining the stability of the polarization measurements. 
Each source has been monitored with 3 aperture sizes (cf. Table~\ref{Table_blob}). 
The SN is circled in black. {The four brightest nearby sources are 
circled in large blue, green, red, and cyan, respectively,} and are labeled with ID numbers also 
used in Table~\ref{Table_blob}. Fainter sources with larger errors are circled in white. 
\label{Fig_image_blob}
}
\end{figure}

For each source and epoch, we measured the flux with three different 
aperture sizes. We used the spreads (full ranges, denoted as `$dq$ range' 
and `$du$ range') in each such set of three measurements to characterize 
their reliability. Bright and highly polarized sources should be less 
affected by noise and hence exhibit a smaller spread, making them useful 
references to check the stability of {\it HST} polarimetry. Because of 
the small number of measurements (three) per source and epoch, which 
renders standard deviations relatively meaningless, we use these spreads 
as proxy of the data quality and instrumental stability. SN\,2014J and 
the four brightest other sources in the field are marked with colored 
circles in Figure~\ref{Fig_image_blob}. Circular apertures of 0$''$.35, 
0$''$.40, and 0$''$.45 were used to measure faint and point sources 
while 0$''$.65, 0$''$.70, and 0$''$.75 were applied for extended sources. 

\begin{figure}[!htb]
\epsscale{0.8}
\plotone{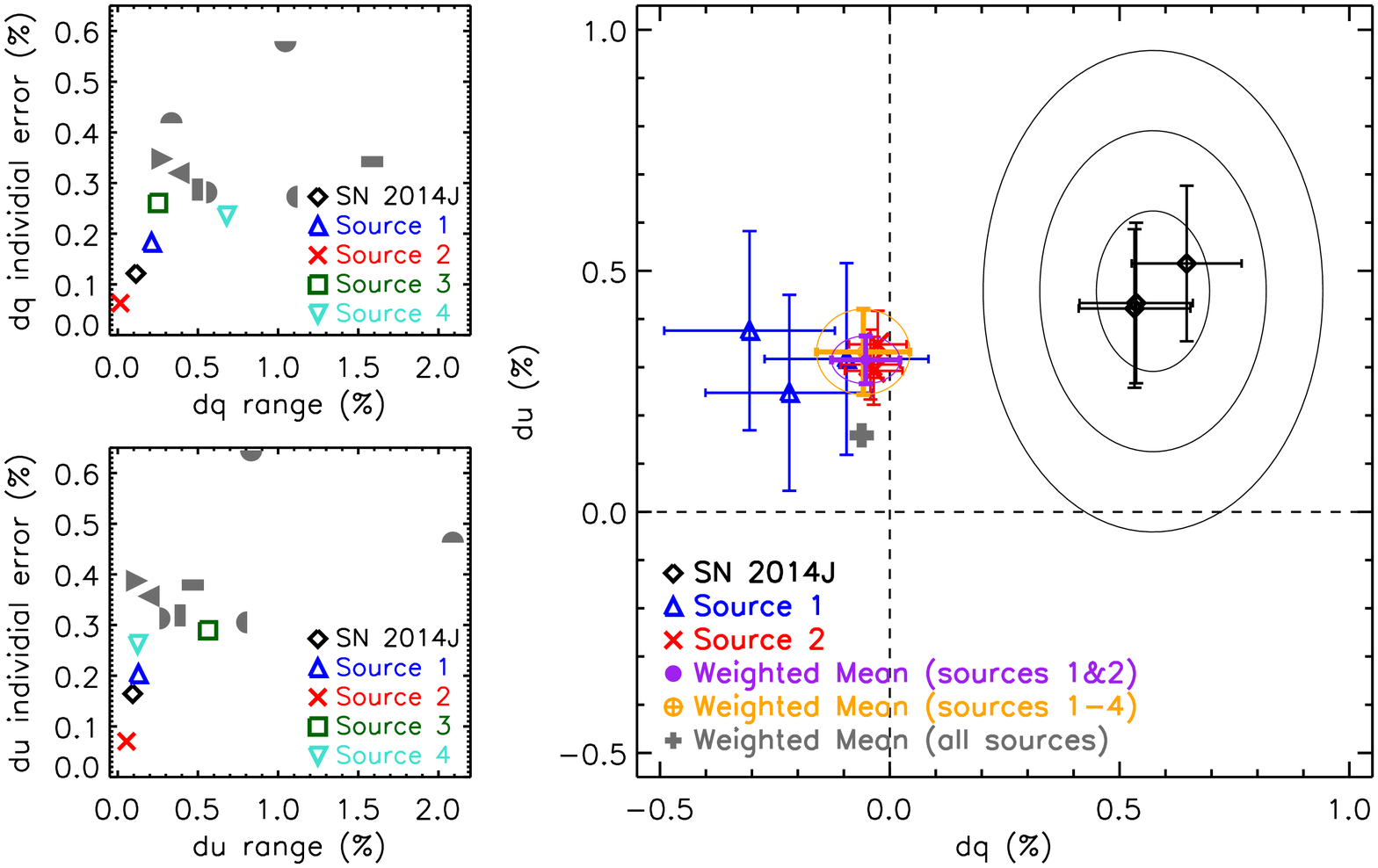}
\caption{\footnotesize Sources used to check the stability of {\it HST} polarimetry. 
{The left panels present for each source the median of the individual 
errors of the polarization measurements with three different apertures as 
a function of spread (ranges) of the measurements with these apertures. 
Similar measurements of other faint field sources are shown by various 
gray symbols, which are the same in the upper and lower panels.} 
In the right panel, the horizontal and vertical axes represent {the 
differences between the $q$ and $u$ values, respectively, measured on 
days 416 and 277.}
SN\,2014J and the {two} brightest other sources (1$-$2) are marked with black and 
colored circles as in Figure~\ref{Fig_image_blob}. 
{The error-weighted mean difference including comparison sources 1 \& 2, 
sources 1$-$4, and all the other fainter sources marked in Figure~\ref{Fig_image_blob} 
are indicated by the purple dot, the orange 
circle with plus sign, and the gray cross, respectively. 
The brightest source, plotted in red, together with the error-weighted mean, 
reveals no time evolution at the {0.3\%} level. Black ellipses show 1, 2, 
and 3-sigma contours centered at the error-weighted mean of SN\,2014J. They 
demonstrate that the variation in polarization of the SN deviates by more 
than 3 times its errors from the variation of the field sources (assumed to 
be intrinsically constant)}. This comparison suggests a genuine evolution of 
the polarization of SN\,2014J between days 277 and 416 (epochs V1 and V2). 
\label{Fig_qu_blob}
}
\end{figure}

In the upper (lower) left panel of Figure~\ref{Fig_qu_blob}, 
{the median of the individual measuring errors in $dq$ ($du$) with 
the three apertures is plotted versus the $dq$ ($du$) range.} 
The sources are identified in Figure~\ref{Fig_image_blob}. 
As shown in Table~\ref{Table_blob}, the individual errors of the three measurements 
of each source are very similar because the apertures differ by only 
$\pm0.05''$ from the median size. Additional measurements of fainter 
field sources appear as various gray symbols; they are the same in the 
two left panels of Figure~\ref{Fig_qu_blob}. The individual errors scale with the 
spread and vice versa. This is expected for well-behaved data. 
Therefore, these graphs confirm the sanity of the data and the method. 
However, the spread is mainly a systematic uncertainty introduced by 
the usage of different apertures while the ordinate illustrates 
photometric errors propergated to the measurement of $dq$ and $du$. 

In the $du$ vs. $dq$ panel of Figure~\ref{Fig_qu_blob}, a significant 
separation of SN\,2014J from the error-weighted mean of all measurements 
on all sources would demonstrate that the polarization of the SN was not 
constant and evolved with time. However, the overall scatter of all 
field sources is dominated by the large errors of the faint sources 
(gray dots). Therefore, we selected those sources whose spreads 
(ranges) in $dq$ and $du$ are less than three times those of 
SN\,2014J. Only sources 1 \& 2 satisfy this criterion. Already for 
the two next fainter sources, 3 \& 4, either the spread in $dq$ or 
$du$ are larger than this limit. 
Other sources were not included because for each of them the spreads 
exceed the threshold in both $dq$ and $du$. 

The measured polarizations of the four brightest comparison sources 
are included in Table 4. The error-weighted mean $dq$ and $du$ values 
of the two brightest field sources were calculated to be 
$\overline{dq}^w = -0.05\pm0.03$\% and 
$\overline{du}^w = 0.32\pm0.04$\%, respectively. 
The results for sources 1$-$4,  
$\overline{dq}^w = -0.06\pm0.03$\% and $\overline{du}^w = 0.33\pm0.04$\% 
are consistent with those for the two brightest sources only. For the 
calculation of the error-weighted means of $dq$, $du$, and their 
associated errors we used the folowing relations: 
\begin{equation}
\overline{x}^w = \frac{\Sigma_{i=i}^{N} x_{i}/\sigma_{i}^2 }{\Sigma_{i=i}^{N} 1/\sigma_{i}^2 }, \  
\sigma_{x}^{w} = \sqrt{\frac{1}{\Sigma_{i=i}^{N} 1/\sigma_{i}^2}} \ , 
\label{Eqn_wmean}
\end{equation}
where $x$ and $\sigma$ denote the measurement and error, respectively, 
of $dq$ or $du$. $N$ is the total number of the individual measurements 
which are numbered by the index $i$. Uncertainties given by 
$\sigma_{x}^{w}$ do not account for the systematic uncertainty 
introduced by the usage of different apertures. We estimated this 
systematic error by calculating the error-weighted standard deviation 
of the individual measurements: 
\begin{equation}
\sigma_{x}^{s} = \sqrt{\frac{\Sigma_{i}^N (x_i - \overline{x}^w)^2 / \sigma_{i}^2}{\frac{N-1}{N} \Sigma_{i=1}^N /\sigma_{i}^2}}, 
\label{Eqn_wstd}
\end{equation}
which gives $\sigma_{dq}^{s}= 0.07$\%, $\sigma_{du}^{s}= 0.03$\% including 
source 1 \& 2, and $\sigma_{dq}^{s}= 0.10$\%, $\sigma_{du}^{s}= 0.08$\% for 
sources 1$-$4. The errors of the error-weighted means shown in Figure~\ref{Fig_qu_blob} were 
estimated by adding $\sigma_{x}^{w}$ and $\sigma_{x}^{s}$ in quadrature. 

The final estimated error-weighted mean and associated error are 
$\overline{dq}^w = -0.05\pm0.07$\% and $\overline{du}^w = 0.32\pm0.05$\% 
based on source 1 \& 2, and $\overline{dq}^w = -0.06\pm0.10$\% and 
$\overline{du}^w = 0.33\pm0.09$\% if sources 3, \& 4 are also included. 
As shown in Figure~\ref{Fig_qu_blob}, the difference in polarization of SN\,2014J between 
day 277 (V1) and day 416 (V2), i.e., $dq = (0.57 \pm 0.12)\%$, $du = (0.46 \pm 0.17) \%$, 
deviates by more than 3 times its error from 
the error-weight mean value calculated from bright sources in the field. 
The error-weighted mean values of $dq$ and $du$ including all the 
marked fainter, and from the previous analysis excluded, sources are 
$dq=-0.06\pm0.03$\% and $du=0.16\pm0.03$\%. That is, unlike SN\,2014J 
there are no general significant systematic differences in polarization 
between epochs V1 and V2. Additionally, the polarization measured in 
different regions of the CCD has previously been shown to agree to 
within 0.2\% \citep{Sparks_etal_2008}. Therefore, we conclude that the 
observed change in polarization of the SN is not an artifact of the 
instrument.} 

\section{Discussion}
{Around optical maximum as well as after day 416, the measured polarizations 
are the same to within the errors but different from those on day 277. 
The deviated degree of polarization on day 277 can be explained by 
light from SN scattered by circumstellar ejecta of $\gtrsim5\times10^{17}$ cm 
($\sim0.5$ light years) from SN~2014J.} 
Compared to the dust detected at day 277, the amount of dust at even 
closer distances from the SN is constrained by the absence, at the 
0.2\% level, of variability of the early polarization. Following 
\citet{Yang_etal_2017a} and the relations between 2-dimensional light 
echoes and 3-dimensional scattering dust distributions 
\citep{Chevalier_1986, Sparks_1994, Sugerman_2003, Tylenda_2004, 
Patat_2005}, we briefly define the geometry of circumstellar light 
echoes used through this paper, 
{also sketched in Figure~\ref{Fig_schem}}. 
The SN is placed at the origin of the plane of 
the sky, a scattering volume element $dV$ lies at distance $r$ from 
the SN, and $z$ gives the foreground distance of the scattering volume 
element along the line of sight. The iso-delay light surface of the 
light echo can be very well approximated by a paraboloid whose focus 
coincides with the SN. We define $\rho$ as the distance from a 
scattering volume element to the SN, projected perpendicular to the 
line of sight (the $z$ direction). The iso-delay light surface gives: 
\begin{equation}
r=\frac{1}{2} \bigg{(} \frac{\rho^2}{ct} + ct \bigg{)}, 
\label{Eqn_echo}
\end{equation}
where $t$ is the time since the SN radiation burst and $c$ denotes the 
speed of light. The scattering angle $\theta$ is therefore given by: 
\begin{equation}
\mathrm{cos} \theta(\rho, t) = z/(z+ct)
\label{Eqn_theta}
\end{equation}

We use the single-clump hypothesis and the scattering angle of $114^\circ$ 
with respect to the line of sight which is implied by the minimum amount of 
astronomical silicate compatible with the observed change in polarization. 
From Equations~\ref{Eqn_echo} and \ref{Eqn_theta}, it then follows that the 
day 33 observations imply less than 
2.6$\times10^{-7} M_{\odot}$ at a distance around 23.5 light days 
(7.3$\times10^{16}$ cm). Similarly, the {\it HST} observations on day 416 
constrain the mass of a single dust clump to less than 
4.0$\times10^{-7} M_{\odot}$ (1 $\sigma$) at a distance around 296 light 
days (7.7$\times10^{17}$ cm). Approximating the radial distribution of the 
clump with a Gaussian function of $\sigma_t=20$ light days generally 
increases the amount of dust by a factor of 2 to 2.5 with respect to the 
above assumed $\delta$ function. A single dust clump is of course an 
over-simplification. The lower limit it places on the mass on day 277 may 
be much larger if the dust is more uniformly distributed, either in a thin 
slab in the plane of the sky at the location of the SN\,2014J or in a more 
radially-extended volume. 

The interpretation of these data is highly model-dependent, but the difference 
of polarization between these epochs and at the SN maximum requires there 
to be either no dust at distances of $\sim 6.1\times10^{16}$ cm (day 33) and 
$\sim 7.7\times10^{17}$ cm (day 416) based on Equation~\ref{Eqn_scatter3} 
(see, i.e., Table~\ref{Table_mass}), or the dust distribution at these 
distances is extremely uniform, such that on the plane of sky the opacity 
fluctuation is less than $\sim$0.002$\pm$0.06 at day 33, and less than 
$\sim$0.0004$\pm$0.0002 at day 416, based on Equation~\ref{Eqn_tau2} and 
assuming $\delta \theta \sim \delta \phi \sim 0.1$. After day $\sim$649, 
the errors of the polarization measurements are much larger but the results 
are still consistent with the polarization at maximum light. Therefore, 
between day $\sim$416 and $\sim$1181, the light from SN\,2014J did not 
encounter significant amounts of dust. 

\subsection{Implications for the Progenitor}
Mass loss through steady stellar wind produces an axially-symmetric ambient 
medium around the line of sight. For an unresolved source, the resultant circumstellar mass profile 
would lead to a cancellation of the vectors of the scattered radiation, 
resulting in zero net circumstellar polarization. Therefore, polarimetry 
cannot independently constrain the mass of material homogeneously 
distributed around a SN. Comprehensive observational studies 
on SN\,2014J disfavor the single-degenerate models with a steady mass 
loss. The absence of a stellar progenitor in pre-explosion images has safely 
ruled out the possibility of red giant donor star \citep{Kelly_etal_2014}. 
The non-detection in X-rays and radio shows a lack of pre-existing material 
to be heated in vicinity of SN\,2014J \citep{Margutti_etal_2014J, 
Perez-Torres_etal_2014}. A combination of numerical models and a late-time 
optical spectrum of SN\,2014J (at 315 days after the explosion) has 
constrained the H-rich unbounded material to be less than 0.0085$M_{\odot}$ 
\citep{Lundqvist_etal_2015}. The {\it Spitzer} mid-infrared observations 
constrain the amount of dust around SN\,2014J to be 
$\lesssim 10^{-5} M_{\odot}$ within a radius of 2.0$\times 10^{17}$ cm 
\citep{Johansson_etal_2017}. 

The gas-to-dust mass ratio for the 8 kpc region around the center of M82 is 
$\sim200$ \citep{Kaneda_etal_2010}. Depending on the dust properties in M82, 
the mass of circumstellar dust clouds depends on the nature and evolution of 
the progenitor system. 
From Table~\ref{Table_mass}, we derive a total minimal mass (dust$+$gas) of 
the CSM responsible for the deviations in late-time degree of polarization 
to be $\gtrsim 5 \times 10^{-4} M_{\odot}$ at a distance of 
$\sim 5.1\times 10^{17}$ cm ($\sim$197 light days), and similar constraints 
on the mass of the CSM are $\sim 2 \times 10^{-5} M_{\odot}$ and 
$\sim 6 \times 10^{-5} M_{\odot}$ at $\sim 6.1\times10^{16}$ cm (day 33) and 
$\sim 7.7\times10^{17}$ cm (day 416), respectively. 
Therein, `CSM' denotes matter lost by the progenitor system whereas `ISM' is 
matter that just happens to be close the location of the progenitor but is 
not related to it. This distinction does not include any a priori implications 
for the distance of such matter from the progenitor. 

\subsubsection{Single-Degenerate Models}
The distance of $\sim$5.1$\times 10^{17}$ cm (197 light days) between 
the dust and the SN can be compared to a putative nova outburst of the 
progenitor prior to the SN explosion. Recurrent nova explosions 
result from a near-Chandrasekhar mass WD accreting at 
$\sim (0.1-3)\times 10^{-7} M_{\odot}yr^{-1}$ and experiencing unsteady 
H burning at its surface (e.g., \citealp{Iben_etal_1982, 
Starrfield_etal_1985, Livio_etal_1992, Yaron_etal_2005}). For a typical 
nova ejection speed of $v_{ej}\sim$1000 km s$^{-1}$, the inferred 
distance between the dust cloud and SN\,2014J is consistent with an 
eruption $t_{ex}\sim$160 years ago. If the nova outburst was brief, the 
ejected mass is likely distributed in a thin clumpy shell. It is also 
possible that the high-speed shell ejection is concurrent with a slower 
wind from the donor star, and any matter surrounding the progenitor was swept up 
by the most recent blast wave (see, i.e., 
\citealp{Wood-Vasey_etal_2006}). These mechanisms in the single-degenerate 
channel can explain the absence of dust closer to and farther away from 
SN\,2014J. We refer to \citet{Margutti_etal_2014J} for a thorough 
discussion on the progenitor configuration for single-degenerate models. 

In some other variants of the single-degenerate model, the SN may have 
exploded inside a planetary nebula shell \citep{Wang_etal_2004, 
Tsebrenko_etal_2013, Tsebrenko_etal_2015}. Numerical models of a type 
Ia SN inside a planetary nebula may explain the observed morphologies 
of the Kepler and G299.2-29 supernova remnants (SNRs, 
\citealp{Tsebrenko_etal_2013}) and G1.9+0.3 SNR 
\citep{Tsebrenko_etal_2015}. The double-shock structure in the 
G1.9+0.3 SNR can be well reproduced by the interaction of type Ia SN 
ejecta with the planetary nebula shell including two or three dense 
clumps \citep{Tsebrenko_etal_2015}. The observed mean radius of the 
G1.9+0.3 SNR is about 2 pc, and in this case, simulations imply the 
total mass in the planetary nebula and the clumps to be 
$\approx 0.09 M_{\odot}$. The size and the total mass of the planetary 
nebula shell vary in different cases, and we consider that the mass 
and the distance of the CSM constrained by our polarimetry of SN\,2014J 
are also broadly consistent with the pre-explosion configuration 
suggested by a SN exploding inside a planetary nebula. 

\subsubsection{Double-Degenerate Models{\label{dd-models}}}
Different double-degenerate models predict different time histories 
for the mass ejection prior to the final explosion triggered by virtue 
of the coalescence between the two WDs. For example, (1) the mass 
stripped and ejected through the `tidal tail' during the dynamics of 
compact WD merger \citep{Raskin_etal_2013}, (2) the mass outflow during 
the unstable final stage of rapid mass accretion immediately preceding 
the merger \citep{Guillochon_etal_2010, Dan_etal_2011}, (3) the outflow 
due to magnetorotationally driven disk wind \citep{Ji_etal_2013}, and 
(4) the ejection of a H-rich layer surrounding a He WD during the 
interaction between a He WD and a C/O WD companion 
\citep{Shen_etal_2013}. 
These four mechanisms predict different masses and locations of dust. 
{\citet{Margutti_etal_2014J} provided a thorough discussion 
based on the {\it Chandra} observation of SN\,2014J. The non-detection 
by {\it Chandra} of CSM around SN\,2014J implies a low-density environment 
with $n_{CSM} \textless 3 \  cm^{-3}$ at $\sim10^{16}$ cm from the SN, 
assuming a wind velocity $v_{ej} \sim$ a few 100 km s$^{-1}$ (or lower) as 
typical velocity of the ejected material.} 
The immediate SN environment depends on $\Delta t_{ex}$, which is the 
time lag between the last major pre-explosion mass ejection and the SN 
explosion. The inferred distance of dust from the SN permits the time 
elapsed since this event to be estimated for an assumed ejection velocity 
($v_{ej}$). In the following, we discuss the predictions of the above 
four mechanisms inferred from the detection of 
$\gtrsim 5 \times 10^{-4} M_{\odot}$ 
CSM at a distance of $\sim 5.1\times 10^{17}$ cm, together with the 
non-detection at a distance of $\sim 6.1\times 10^{16}$ cm and beyond 
the distance of $\sim 7.7\times10^{17}$ cm. 

(1) Tidal tail ejection. Prior to coalescence, as a major consequence 
of a merger of two compact WDs, a small fraction of the system mass 
will be expelled and leave the system at the escape velocity. 
A 3D hydrodynamics simulation shows that a mass of 
$(1-5) \times 10^{-3} M_{\odot}$ will be lost from the system and achieve an escape 
velocity of $v_{ej} \approx 2,000$ km s$^{-1}$ \citep{Raskin_etal_2013}. 
The ejecta are highly nonaxisymmetric and have opening angles of 
$\approx 93^{\circ}$ and $\approx 41^{\circ}$ in the plane of the 
disk and perpendicular to it, respectively. 
The estimated mass and 
inferred clumpy profile of the ejecta from our observations of SN\,2014J 
both agree well with the predictions of tidal tail ejection 
\citep{Raskin_etal_2013}. However, the time lag, $\Delta t_{ex}$, which 
determines the distance of the pre-explosion ejecta is unclear. 
For $v_{ej} \approx 2,000$ km s$^{-1}$, our observations indicate 
that $\Delta t_{ex} \approx 80$ years. 

(2) Mass outflows during rapid accretion. \citet{Guillochon_etal_2010} 
and \citet{Dan_etal_2011} have shown that the mass transfer between a 
pure He WD or a He/CO hybrid and a CO WD can be 
unstable. Therefore, high-density regions may build up that lead to surface 
detonations that trigger the final thermonuclear runaway. During the 
rapid mass accretion process (with rates reaching 
$\sim10^{-5} - 10^{-3} M_{\odot} \ s^{-1}$ at final tens of orbits), a mass 
of $M_{ej}\sim 10^{-2} - 10^{-3} M_{\odot}$ will be lost through the 
system's Roche surface at $v_{ej}$ $\sim$ a few 1,000 km s$^{-1}$. Our 
observations would imply that substantial material can be ejected as early 
as several tens of years before the coalescence. This is comparable 
to the mass limit at $\sim10^{16}$ cm set by the {\it Chandra} X-ray 
observation \citep{Margutti_etal_2014J}. 

(3) Disk winds. During the WD-WD merger, an unstable 
magnetorotationally-driven accretion disk will be produced before the 
detonation leading to the explosion of a type Ia SN. 
Simulations suggest that about $10^{-3}M_{\odot}$ will become 
gravitationally unbound, and being ejected at a mean velocity 
$v_{ej}$ $\sim$2600 km s$^{-1}$ \citep{Ji_etal_2013}. This outflow 
produced by magnetorotationally driven turbulence within the disk 
yields a similar time history of the mass ejection as that predicted 
by the tidal tail ejection \citep{Raskin_etal_2013}. Our observations 
suggest $\Delta t_{ex} \approx 60$ years. These magnetized outflows 
are predicted to be strongly nonaxisymmetric, with an opening angle 
of $\sim 50^{\circ}$. This is also consistent with 
the inferred clumpy structure of the circumstellar dust cloud. 

(4) Shell ejection. In a system with a C/O WD accreting He from 
a He-burning star, an explosion in the He layer would trigger the 
detonation of the C/O core \citep{Livne_etal_1990}. In this 
double-detonation context, \citet{Shen_etal_2013} have proposed that a 
H-rich layer surrounding the He core WD would impact the mass transfer 
and its ejection. Their simulations suggest that the H-rich material 
will be removed from the binary system through multiple mass ejections 
over the course of 200-1400 years prior to the merger. The total 
ejected mass is $M_{ej} = (3-6) \times 10^{-5} M_{\odot}$ and 
$v_{ej} \approx 1500$ km s$^{-1}$, roughly equal to 
the velocity of a He WD in a circular orbit. 
Our polarimetric tomography of the circumstellar 
environment around SN\,2014J did not find significant amount of dust at 
distance\footnote[3]{Assuming the scattering angle of the dust cloud 
to be 90$^{\circ}$} of $\sim 8\times10^{16}$ (day 33), and beyond 
$\sim(1-3)\times 10^{18}$ cm (after day 416). The detected 
$\gtrsim 5 \times 10^{-4} M_{\odot}$ CSM at a distance of 
$\sim 5.1\times 10^{17}$ cm is roughly an order of magnitude larger 
than the total mass predicted 
for shell ejection. Furthermore, $\Delta t_{ej} \approx 107$ year is 
less than the 200-1400 years expected from the model. 

Based on the above interpretations, we conclude that the mass of the 
pre-explosion ejecta and the time delay between such an event 
and the SN explosion are broadly consistent with most of the 
double-degenerate models discussed in \citet{Margutti_etal_2014J}. 
While the polarimetry of SN\,2014J contributes important information 
for the understanding of the nature and pre-explosion evolution of 
the progenitors of type Ia SNe, it cannot discriminate between 
single- and double-degenerate models. 
We are also unsure about whether the double-degenerate models 
provide the proper temperature and density on the right timescale 
to enable dust formation in the implied time, i.e., several to ten 
decades. This issue needs to be addressed in the future. 

\subsection{Polarization Position Angles and Dust Alignment}
{ 
As shown in Table~\ref{Table_pol}, the degree of polarization 
decreased to $\sim3.8$\% on day 277 from the interstellar polarization ($\sim4.9$\%), 
and restored to $\sim4.6$\% on day 416.} However, 
the polarization position angles at day 277 and day 416 exhibit no 
time evolution, except in the $i$-band data taken on day 416, where 
the degree of polarization is low and the $PA$ suffers larger 
uncertainties. A possible explanation is that the dust particles in 
the scattering cloud(s) are nonaxisymmetric and aligned with the 
foreground dust that is responsible for the extinction. 
The magnetic field close to the SN progenitor may be 
highly coherent and very efficient in quickly aligning dust particles. 
This is qualitatively discussed in the following paragraphs.

\begin{figure}[!htb]
\epsscale{1.0}
\plotone{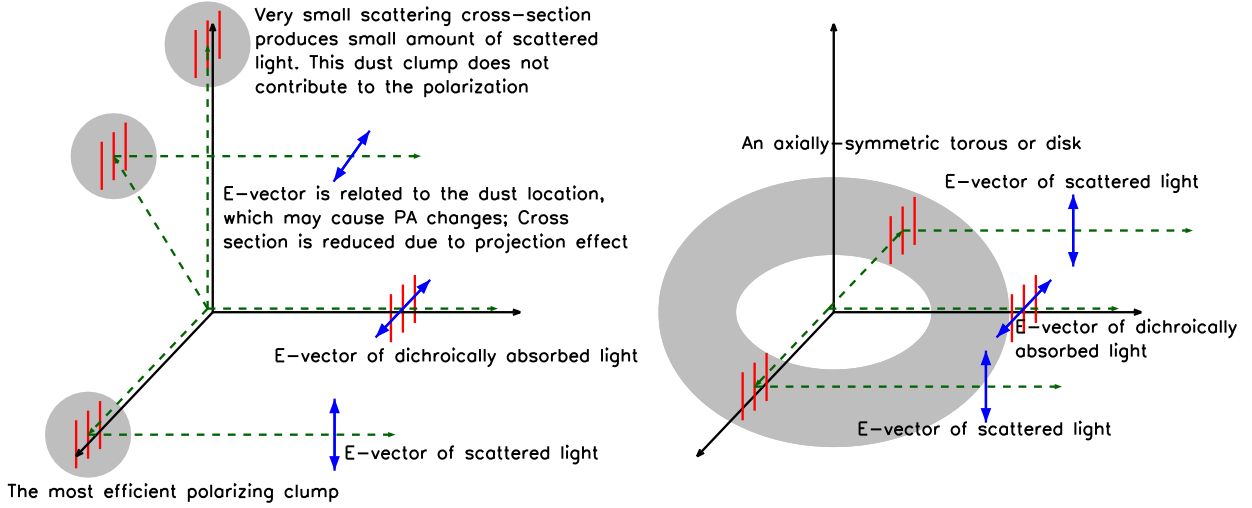}
\caption{\footnotesize Schematic diagram explaining the 
non-local coherence of the polarization $PA$ in the case that 
the grains in circumstellar dust clumps are aligned with 
the local interstellar magnetic field. Red bars illustrate dust 
grains aligned by an ad-hoc coherent magnetic field, green dashed 
lines represent light from the SN, blue arrows demonstrate the 
direction of E-vectors of the net polarized light. The observer is 
located outside the right edge of the figure. {In the right panel, 
the net effect is a rotation in the $QU$ plane through $180^{\circ}$ 
therefore the scattered light does not impose a rotation on the 
$PA$ of integrated light measured from the SN point source.} 
\label{Fig_align}
}
\end{figure}

When light from a SN is scattered by circumstellar dust grains, the 
E-vector will be perpendicular to the scattering plane so that the 
polarization $PA$ is related to the location of the dust, here 
approximated by a single clump. We also assume that the cross-section 
of aligned dust grains is larger along their major axis, and the 
polarization is strongest when the grains' major axis is perpendicular 
to the scattering plane. For instance, needle-like grains at a right 
angle to the scattering plane can produce a significant amount of 
polarization. If a large-scale magnetic field permeates both the 
circumstellar dust and the line-of-sight ISM, it may align the grains 
in the dust clump and in the ISM to the same direction. Consequently, 
the E-vector of dichroically-absorbed light on the direct SN-Earth 
line of sight is normal to that of the light scattered by circumstellar 
dust. Figure~\ref{Fig_align} gives schematic views of the net E-vector 
generated by circumstellar scattering and dichroic extinction. 

The scattered light will be polarized with the E-vector perpendicular 
to the scattering plane, whereas the transmitted light will have an 
E-vector preferably absorbed in this direction. If the dust grains in 
the foreground ISM and the circumstellar dust are both aligned by the 
same local interstellar magnetic field, this explains why the 
polarization decreases as the unresolved circumstellar light echo 
studied in this paper emerges. Most efficiently scattering (and 
polarizing) dust consists of particles aligned with the ambient 
magnetic field. Under this assumption, the aligned interstellar grains 
do not impose a rotation on the integrated polarization of the SN point 
source. Otherwise, the scattered light may contribute only a few percent 
to the total received light so that the rotation is small (i.e. barely 
measurable). This holds even in the more general case in which the 
scattering polarization in the resolved circumstellar light echoes and 
the direct line-of-sight interstellar polarization are not perpendicular. 
However, if the circumstellar light echoes are contributing more 
substantially to the total signal, rotation in the integrated $PA$ with 
respect to the interstellar direction is expected if the polarization 
$PA$ in circumstellar light echoes is not perpendicular to the local 
interstellar magnetic field. 

This reasoning permits an independent limit to be set on the flux 
contribution of the light echo. The observed polarization is a vector 
combination of the interstellar polarization and circumstellar polarization. 
After correcting the instrumental polarization and projecting these two 
polarization components on the $Q$ and $U$ axes, we can rewrite 
Equations~\ref{Eqn_p} and \ref{Eqn_pa} as follows: 
\begin{equation}
p\% = \sqrt{(q_{isp} + q_{csp})^2 + (u_{isp}+u_{csp})^2},
\label{Eqn_p1}
\end{equation}
\begin{equation}
PA = \frac{1}{2} \mathrm{tan^{-1}} \bigg{(} \frac{u_{isp} + u_{csp}}{q_{isp}+q_{csp}} \bigg{)},
\label{Eqn_pa1}
\end{equation}
where superscripts ${isp}$ and ${csp}$ correspondingly denote the 
interstellar and circumstellar polarization. 
{If we assume the polarization imparted by the scattering is $\sim$50\%, 
Equation~\ref{Eqn_p1} and \ref{Eqn_pa1} imply that, if the maximal 
change in $PA$ is $\sim3^{\circ}$, the polarized flux contributed by a 
light echo to the total polarized flux observed from SN\,2014J should 
not exceed $\sim$10\%, and the contribution by a light echo to the total 
observed flux from SN\,2014J should not exceed 1\%. 
This $\sim3^{\circ}$ variation in $PA$ is comparable to the observed 
$\Delta PA = 2^{\circ}.6 \pm 1^{\circ}.0$ in $F475W$ from V1 and V2.} 
For the most efficient case of circumstellar polarization, i.e., 
by a single clump of astronomical silicate with $a=0.1 \ \mu m$ at 
$\sim$114$^{\circ}$ (Section~\ref{Section_model}), the polarization decrease 
observed on day 277 (from $\sim$4.9\% to $\sim$3.8\%) can be explained with 
a $\sim$1\% flux contribution from the light echoes in the $F475W$-band as 
is also illustrated by the inset panel in Figure~\ref{Fig_pol_time}. 
{Therefore, based on the deviant integrated degree of polarization and the 
invariant $PA$ observed on day 277, 
we infer that the rotation of the $PA$ 
introduced by the circumstellar light echoes around SN\,2014J is less than 
$\sim3^{\circ}$ with respect to the interstellar polarization.} This number 
is model dependent, but the most efficient configuration for producing 
polarized light is that in which the circumstellar dust grains are aligned 
with the ambient interstellar magnetic field. In this scheme, we discuss 
some other implications as follows. 

Circumstellar dust composed of needle-like grains aligned with the 
interstellar magnetic field has a net polarizing effect even if its 
spatial distribution is spherically symmetric. The reason is that 
scattering in planes aligned with the grains would produce zero 
polarization. Therefore, it would not lead to a cancellation of the 
polarization produced by scattering on planes perpendicular to the 
dust alignment, and a net polarization arises (as illustrated 
by Figure~\ref{Fig_schem}). 
This indicates that the polarization of light echoes is not necessarily 
an indication of the nonaxisymmetry of the dust distribution. 

Polarization traces the magnetic field and enables a unique approach 
to the study of its interaction with nonaxisymmetric dust. Careful 
studies of dust grains aligned through the `radiative alignment 
torque' (RAT) are able to provide testable predictions on various 
properties \citep{Lazarian_etal_2007}. 
\citet{Andersson_etal_2010} found that dust surrounding the Herbig 
Ae/Be star HD 97300 does not align with the stellar wind, ruling out 
significant contributions to grain alignment through the stellar wind 
or radiation pressure of the star (the so-called Gold alignment, see, 
i.e., \citealp{Gold_etal_1952}). At a star-cloud distance of 
$\sim$0.03 pc, a weak dependence of the grain alignment efficiency on 
the angle between the magnetic field and the radiation field 
anisotropy is seen \citep{Andersson_etal_2010}. This confirms the 
grain alignment predicted by the RAT model. Furthermore, dust grains 
in the vicinity of a type Ia SN may be more efficiently aligned by 
the radiative torque of the SN radiation. Among all the single- and 
double-degenerate cases, 
with time lags around tens to a hundred years, the growth in grain size 
of the pre-explosion ejecta may not be particularly relevant 
considering the relatively long time-scale of the grain 
growth, see Figure 8 of \citet{Mattsson_2016}. 
These small grains can be effectively aligned by the SN radiation 
regardless of the relatively small effect of the interstellar 
radiation field. For instance, 
at distances of 1$-$10 pc, $a\sim0.03 \ \mu m$ grains can be radiatively 
aligned within $\sim 0.5-40$ days for SN luminousity of $10^{8} L_{\odot}$ 
\citep{Hoang_etal_2017}. However, lacking further observational constraints, 
we conclude that it still remains unclear, 
whether an intrinsic magnetic field of the progenitor of 
SN\,2014J or the ambient magnetic field in the ISM of M82 could align 
the dust grains quickly enough within the relatively short time 
(estimated above at $\sim$160 years) between the pre-explosion mass 
ejection and the SN explosion. 

It is also possible that the dust grains in the pre-explosion ejecta 
are aligned neither with a magnetic field nor the radiation torque of 
the SN radiation, i.e., exhibit no dominant directional preference. 
Instead of being elongated but randomly 
oriented, dust grains may alternatively have nearly spherical shape 
with little polarizing power because the difference between minimal 
and maximal extinction efficiencies is small. In all these cases, 
the deviant integrated degree of polarization and the invariant $PA$ 
observed on day 277 require the dust to be at certain position angles 
relative to the SN, i.e., the scattering plane is perpendicular to 
the interstellar magnetic field. This would introduce an orthogonal 
polarization component to the integrated light. Under these 
circumstances, the vectorial combination of the two components only 
affects the degree of polarization but not the $PA$ as observed in 
SN\,2014J.

It is important to stress that resolved light echoes around SN\,2014J 
caused by interstellar dust \citep{Crotts_2015, Yang_etal_2017a} do 
not compromise the inference of circumstellar dust from the evolution 
of non-spatially-resolved polarization. The scattering angle by 
foreground ISM is $\theta \sim \sqrt{2ct/z} \sim 4.5^{\circ} 
(\frac{t}{1year} \frac{100 pc}{z})^{1/2}$, where $t$ denotes the time 
after optical maximum and $z$ is the foreground distance of the dust 
to the SN. At such small scattering angles, the polarization of resolved 
light echoes results from the dichroic extinction by partially aligned 
non-spherical paramagnetic dust grains. This interstellar polarization 
can be determined from the SN polarization around maximum light 
\citep{Kawabata_etal_2014, Patat_etal_2015}. Moreover, any such 
polarization signal that at the distance of M82 is unresolved by 
{\it HST} is expected to be constant with time. Therefore, it cannot 
explain the deviant measurement on day 277. 


One other possibility is that SN\,2014J exploded close to some pre-existing 
interstellar dust clouds. The morphological evolution of the `luminous arc' 
light echo probed by the iso-delay light surface at day 277 and day 416 
reveals the inhomogeneity of the foreground ISM 
which transformed from three clumps to two short segments of concentric 
arcs (see, Figure 4 of \citealp{Yang_etal_2017a}). This implies that the 
ISM in the vicinity of the SN\,2014J-Earth line of sight is inhomogeneous 
on scales smaller than $\sim$2.3 pc at a foreground distance of 226 pc 
(see the projected radius at day 277 and day 416 in 
Table 4 of \citealp{Yang_etal_2017a}). 
{This does not invalidate the claims of the small $R_V$ variations of 
Galactic dust in a local kilo-parsec volume probed with a spatial resolution 
of $\sim$60 pc and within only $\sim$100 pc scale height ($R_V=3.0\pm0.2$, 
\citealp{Schlafly_etal_2017}). Recently, based on low-resolution 
spectro-polarimetric observations of multiple sight-lines, 
\citet{Siebenmorgen_etal_2017} found significant variations of the Galactic 
dust characteristics on small scales, and from cloud-to-cloud (i.e., 
$2.3 \leq R_V \leq 5.0$). Smaller scales of inhomogeneity of the ISM in M82 
can still be possible.} 
Therefore, we cannot rule out the possibility that the scattering dust 
cloud(s) producing the late-time deviation in polarimetry of SN\,2014J is 
part of the ISM close to the SN\,2014J. 

\section{Summary}
Monitoring with the imaging polarimetry mode of the {\it HST} ACS/WFC 
at six epochs from 277 and 1181 days after the maximum light has probed 
the circumstellar environment of the type Ia supernova 2014J. The 
polarization exhibited a conspicuous deviation on day 277 from all 
other epochs. This difference can result from light scattered by 
circumstellar ejecta of $\gtrsim 5 \times 10^{-4} M_{\odot}$ located 
$\sim 5\times 10^{17}$ cm ($\sim$0.5 light years) from SN\,2014J. 
The polarization at other epochs is consistent with the interstellar 
polarization around the optical maximum.  This rules out significant 
circumstellar dust at distances between $\sim$1 light year and 
$\sim$3.3 light years from SN\,2014J. If attributed to the progenitor 
of SN\,2014J, the distance of the dust from the SN constrains the time 
of ejection. It is consistent with a single-degenerate model with an 
unsteady mass loss, i.e., experiencing a nova-like eruption about 160 
years before the SN explosion for a typical speed of 1,000 km s$^{-1}$. 
The inferred mass and distance of the circumstellar dust cloud 
are also consistent with an explosion inside a planetary nebula 
including dense clumps. 

In most of the double-degenerate models, a significant amount of mass 
($\sim 10^{-4} - 10^{-2} M_{\odot}$) will be ejected prior to the 
coalescence between the two WDs. {The time lag between the pre-explosion 
mass ejection and the final explosion ranges from hundreds of seconds 
to $\sim100$ years, depending on the model we have discussed in 
Section~{\ref{dd-models}}.} The mass loss history 
deduced from the late-time polarimetry of SN\,2014J is consistent with 
most of the double-degenerate scenarios discussed in 
\citet{Margutti_etal_2014J} and references therein. 
In spite of providing important constraints on the nature and pre-explosion 
evolution of the progenitors of type Ia SNe, our time-resolved precision 
polarimetry with {\it HST} could not 
discriminate between single- and double-degenerate models. 

The single-event-like time dependence of the degree of the 
polarization and the constancy of the polarization angle can be 
understood if the circumstellar dust of SN\,2014J is aligned with the 
ambient interstellar magnetic field. {However, both grains with low 
asymmetry and elongated grains aligned by the radiative torque by the 
progenitor's radiation could lead to the same effect if the dust cloud 
is located at an angle of $\sim90^{\circ}$ to the position angle of the 
ambient interstellar polarization.}  
Polarimetry of light echoes around Galactic novae can enable critical 
tests of the alignment mechanism of dust grains. 

We have presented a novel method for probing the circumstellar environment 
of type Ia SN. This method uses the time evolution of SN polarization at 
late epochs to constrain the mass and distance of material inhomogeneously 
distributed around the SN. When a significant time evolution of 
polarization is observed at a location close to the SN as implied by the 
elapsed time and the angular separation, we will be able to place stringent 
constraints on the presence of circumstellar dust. Although our current 
data cannot place solid criteria to distinguish between the single- and 
double-degenerate channels for type Ia SNe explosion, polarimetry 
at late times may emerge as a new and effective way of systematically 
studying the progenitor systems of type Ia SNe. Future observations of the 
type Ia SNe at late epochs will help to address the nature of circumstellar 
dust around type Ia SNe and their effect on the reddening and extinction 
towards the SNe. 

\acknowledgments 
We greatly appreciate Dave Borncamp and the {\it HST} ACS team in 
fixing the distortion correction issues in ACS/WFC polarized images. 
We would like to thank the anonymous referee for the very careful scrutiny 
which resulted in very helpful, constructive suggestions that improved the paper. 
We also thank Jian Gao, Bi-wei Jiang, Kevin Krisciunas, Armin Rest, 
and Nicholas Suntzeff for helpful discussions. Some of the data used 
in this study were obtained from the Mikulski Archive for Space 
Telescopes (MAST). PyRAF, PyFITS, STSCI$\_$PYTHON are products of the 
Space Telescope Science Institute, which is operated by AURA for NASA.
STScI is operated by the Association of Universities 
for Research in Astronomy, Inc., under NASA contract NAS5-26555. 
Support for MAST for non-HST data is provided by the NASA Office of 
Space Science via grant NNX09AF08G and by other grants and contracts.
The supernova research by Y. Yang,
P. J. Brown, and L. Wang is supported by NSF grant AST-0708873.
P. J. Brown was partially supported by a Mitchell Postdoctoral
Fellowship.  Y. Yang and M. Cracraft also acknowledge support from
NASA/STScI through grant HST-GO-13717.001-A, grant HST-GO-13717.001-A, 
HST-GO-14139.001-A, and HST-GO-14663.001-A. 
{The research of Y. Yang 
is also supported through a Benoziyo Prize Postdoctoral Fellowship.} 
The research of J. Maund 
is supported through a Royal Society University Research Fellowship.
L. Wang is supported by
the Strategic Priority Research Program ``The Emergence of Cosmological
Structures'' of the Chinese Academy of Sciences, Grant No. XDB09000000.
L. Wang and X. Wang are supported by the Major State Basic Research
Development Program (2013CB834903), and X. Wang is also supported by
the National Natural Science Foundation of China (NSFC grants 11178003, 
11325313, and 11633002). 

\begin{deluxetable}{ccccccccccc}
\tablewidth{0pc}
\tabletypesize{\scriptsize}
\tablecaption{Log of observations of SN\,2014J with $HST$ ACS/WFC $POL*V$ \label{Table_log}}
\tablehead{
\colhead{Filter} \vspace{-0.0cm}  & \colhead{Polarizer}  & \colhead{Date}  & \colhead{Exp}  & \colhead{Phase$^a$} & \colhead{Date} & \colhead{Exp} & \colhead{Phase$^a$} & \colhead{Date} & \colhead{Exp} & \colhead{Phase$^a$} \\
\colhead{} &  &  \colhead{(UT)} & \colhead{(s)}  & \colhead{(Days)} & \colhead{(UT)} & \colhead{(s)} & \colhead{(Days)} & \colhead{(UT)} & \colhead{(s)} & \colhead{(Days)}  }
\startdata
F475W & POL0V   & 2014-11-06 &  3$\times$130  & 276.5   & 2015-03-25 &  3$\times$400  & 415.6  & 2015-11-12 &  4$\times$1040 & 648.5  \\
F475W & POL120V & 2014-11-06 &  3$\times$130  & 276.5   & 2015-03-25 &  3$\times$400  & 415.6  & 2015-11-12 &  4$\times$1040 & 648.7  \\
F475W & POL60V  & 2014-11-06 &  3$\times$130  & 276.5   & 2015-03-25 &  3$\times$400  & 415.7  & 2015-11-12 &  4$\times$1040 & 648.8  \\
F606W & POL0V   & 2014-11-06 &  2$\times$40   & 276.6   & 2015-03-27 &  3$\times$60   & 417.9  & 2015-11-12 &  4$\times$311  & 649.0  \\
F606W & POL120V & 2014-11-06 &  2$\times$40   & 276.6   & 2015-03-27 &  3$\times$60   & 418.0  & 2015-11-13 &  4$\times$311  & 649.0  \\
F606W & POL60V  & 2014-11-06 &  2$\times$40   & 276.6   & 2015-03-27 &  3$\times$60   & 418.0  & 2015-11-13 &  4$\times$311  & 649.1  \\
F775W & POL0V   & 2014-11-06 &  2$\times$30   & 276.6   & 2015-03-27 &  3$\times$20   & 418.0  & 2015-11-12 &  4$\times$100  & 648.5  \\
F775W & POL120V & 2014-11-06 &  1$\times$55   & 276.6   & 2015-03-27 &  3$\times$20   & 418.0  & 2015-11-12 &  4$\times$100  & 648.7  \\
F775W & POL60V  & 2014-11-06 &  1$\times$55   & 276.6   & 2015-03-27 &  3$\times$20   & 418.0  & 2015-11-12 &  4$\times$100  & 648.9  \\
                \hline
F475W & POL0V   & 2016-04-08 &  4$\times$1040 & 796.2   & 2016-10-12 &  4$\times$1040 & 983.1  & 2017-04-28 &  4$\times$1040 & 1181.3 \\
F475W & POL120V & 2016-04-08 &  4$\times$1040 & 796.4   & 2016-10-12 &  4$\times$1040 & 983.3  & 2017-04-28 &  4$\times$1040 & 1181.4 \\
F475W & POL60V  & 2016-04-08 &  4$\times$1040 & 796.6   & 2016-10-12 &  4$\times$1040 & 983.4  & 2017-04-28 &  4$\times$1040 & 1181.5 \\
F606W & POL0V   & 2016-04-08 &  4$\times$311  & 796.8   & 2016-10-14 &  3$\times$360  & 985.1  & 2017-04-28 &  3$\times$360  & 1181.7 \\
F606W & POL120V & 2016-04-08 &  4$\times$311  & 796.8   & 2016-10-14 &  3$\times$360  & 985.1  & 2017-04-28 &  3$\times$360  & 1181.7 \\
F606W & POL60V  & 2016-04-08 &  4$\times$311  & 796.9   & 2016-10-14 &  3$\times$360  & 985.1  & 2017-04-28 &  3$\times$360  & 1181.7 \\
F775W & POL0V   & 2016-04-08 &  4$\times$100  & 796.2   & 2016-10-12 &  4$\times$202  & 983.1  & 2017-04-28 &  4$\times$202  & 1181.3 \\
F775W & POL120V & 2016-04-08 &  4$\times$100  & 796.4   & 2016-10-12 &  4$\times$202  & 983.3  & 2017-04-28 &  4$\times$202  & 1181.4 \\
F775W & POL60V  & 2016-04-08 &  4$\times$100  & 796.6   & 2016-10-12 &  4$\times$202  & 983.4  & 2017-04-28 &  4$\times$202  & 1181.5 \\
\enddata
\tablenotetext{a}{Days since B maximum on 2014 Feb. 2.0 (JD 245 6690.5).}
\label{Table_log}
\end{deluxetable}

\begin{table}[!htb]
\caption{The polarization degree of SN\,2014J \label{Table_pol}}
\begin{scriptsize}
\begin{tabular}{ccccc|cccc}
\hline
Filter  & Phase  &   $p$  & $PA$    & mag & Phase          & $p$ & $PA$       & mag\\
        & {Days} & $\%$ & degrees &     & {Days}         & $\%$ & degrees &    \\
\hline
$F475W$ & 276.5 & 3.82$\pm$0.12 & 40.3$\pm$0.9 & 17.363$\pm$0.001 & 415.6 & 4.56$\pm$0.21 & 37.7$\pm$1.2 & 19.464$\pm$0.002 \\
$F606W$ & 276.6 & 2.65$\pm$0.21 & 46.9$\pm$2.3 & 17.429$\pm$0.002 & 417.9 & 3.27$\pm$0.48 & 43.4$\pm$3.5 & 19.594$\pm$0.003 \\
$F775W$ & 276.6 & 1.19$\pm$0.24 & 41.7$\pm$7.5 & 16.742$\pm$0.002 & 418.0 & 1.55$\pm$0.58 & 17.1$\pm$6.2 & 18.268$\pm$0.004 \\
\hline
$F475W$ & 648.5 & 4.68$\pm$0.44 & 33.3$\pm$2.6 & 22.363$\pm$0.003 & 796.2 & 3.50$\pm$0.81 & 33.0$\pm$6.6  & 23.266$\pm$0.006 \\
$F606W$ & 649.0 & 4.57$\pm$0.58 & 47.7$\pm$3.7 & 21.962$\pm$0.005 & 796.8 & 0.78$\pm$1.19 & 73.2$\pm$43.6 & 22.917$\pm$0.009 \\
$F775W$ & 648.5 & 4.49$\pm$0.75 & 39.9$\pm$4.8 & 21.427$\pm$0.006 & 796.2 & 2.40$\pm$1.48 & 54.1$\pm$17.5 & 22.492$\pm$0.011 \\
\hline
$F475W$ & 983.1 & 2.27$\pm$1.84 & 48.3$\pm$23.6 & 24.169$\pm$0.014 & 1181.4 & 5.61$\pm$2.76 & 59.2$\pm$16.0 & 24.765$\pm$0.023 \\
$F606W$ & 985.1 & 6.58$\pm$3.09 & 53.5$\pm$13.9 & 23.934$\pm$0.024 & 1181.7 & 3.12$\pm$5.88 & 37.4$\pm$53.2 & 24.695$\pm$0.049 \\
$F775W$ & 983.1 & 8.43$\pm$1.99 & 68.3$\pm$6.8  & 23.294$\pm$0.015 & 1181.4 & 7.61$\pm$4.19 & 104.6$\pm$15.5 & 24.234$\pm$0.032 \\
\end{tabular}
\end{scriptsize}
\end{table}

\begin{table}[!htb]\centering \caption{Minimal dust masses implied by the observed polarization
\label{Table_mass}}
\begin{scriptsize}
\begin{tabular}{c c c c c c}\hline\hline
\multicolumn{1}{c}{\textbf{Epoch}} & \textbf{Dust}
 & \textbf{$\theta_{max}$} & \textbf{r} &  \textbf{Mass$(\theta_{max})$} &  \textbf{Mass$(\theta_{90^{\circ}})$} \\
\textbf{(Days)} & & \textbf{($^\circ$)} & \textbf{(cm)} & \textbf{($M_{\odot}$)} & \textbf{($M_{\odot}$)} \\ \hline
      & Milky Way & 100$_{-4}^{+4}$ & $7.3_{-0.4}^{+0.5}\times 10^{16} $ & $(-0.1 \pm 2.7) \times10^{-6} $ & $(-0.1 \pm 2.8) \times 10^{-6} $ \\
t=33$^1$  & Silicate  & 114$_{-5}^{+5}$ & $6.1_{-0.3}^{+0.4}\times 10^{16} $ & $(-0.1 \pm 2.1) \times10^{-7} $ & $(-0.1 \pm 3.2) \times 10^{-7} $ \\
      & Graphite  &  92$_{-5}^{+5}$ & $8.3_{-0.6}^{+0.7}\times 10^{16} $ & $(-0.1 \pm 3.0) \times10^{-7} $ & $(-0.1 \pm 3.0) \times 10^{-7} $ \\
\hline     
      & Milky Way & 100$_{-4}^{+4}$ & $6.1_{-0.3}^{+0.4}\times 10^{17} $ & $(3.2 \pm 0.4)  \times10^{-5} $ & $(3.6 \pm 0.4)  \times 10^{-5} $ \\
t=277 & Silicate  & 114$_{-5}^{+5}$ & $5.1_{-0.3}^{+0.3}\times 10^{17} $ & $(2.5 \pm 0.3)  \times10^{-6} $ & $(3.7 \pm 0.4)  \times 10^{-6} $ \\
      & Graphite  &  92$_{-5}^{+5}$ & $6.9_{-0.5}^{+0.6}\times 10^{17} $ & $(3.6 \pm 0.4)  \times10^{-6} $ & $(3.6 \pm 0.4)  \times 10^{-6} $ \\
\hline     
      & Milky Way & 100$_{-4}^{+4}$ & $9.2_{-0.5}^{+0.6}\times 10^{17} $ & $(3.7 \pm 1.9)  \times10^{-6} $ & $(4.1 \pm 2.1)  \times 10^{-6} $ \\
t=416 & Silicate  & 114$_{-5}^{+5}$ & $7.7_{-0.4}^{+0.5}\times 10^{17} $ & $(2.9 \pm 1.4)  \times10^{-7} $ & $(4.3 \pm 2.2)  \times 10^{-7} $ \\
      & Graphite  &  92$_{-5}^{+5}$ & $1.0_{-0.1}^{+0.1}\times 10^{18} $ & $(4.1 \pm 2.1)  \times10^{-7} $ & $(4.2 \pm 2.1)  \times 10^{-7} $ \\  
\hline
\end{tabular}\\
{$^1$}{The negative masses on day 33 are due to an opposite sign of the differences from the interstellar 
foreground polarization compared to day 277 and day 416. }
\end{scriptsize}
\end{table}


\begin{table}[!htb]
\caption{Polarizations of other bright sources in the {\it HST} ACS/WFC field\label{Table_blob}}
\begin{scriptsize}
\begin{tabular}{cccccccccc}
\hline
 & R.A.(J2000) & Dec (J2000) & Aperture                 & $q^1$  & $q^2$  & $u^1$  & $u^2$   & $I^1$       & $I^2$  \\
 & (h:m:s)     & (d:m:s)     & (radius in $\arcsec$)    & ($\%$) & ($\%$) & ($\%$) & ($\%$)  & (counts/s)  & (counts/s)\\
\hline
         &             &             & 0.35 & -0.61$\pm$0.08 & -1.15$\pm$0.12 & -3.79$\pm$0.08 & -4.22$\pm$0.12 & 2695.0$\pm$1.6 & 399.3$\pm$0.3 \\
SN 2014J & 09:55:42.11 & 69:40:25.90 & 0.40 & -0.60$\pm$0.08 & -1.13$\pm$0.12 & -3.82$\pm$0.08 & -4.24$\pm$0.12 & 2736.6$\pm$1.6 & 415.2$\pm$0.3 \\
         &             &             & 0.45 & -0.57$\pm$0.08 & -1.22$\pm$0.12 & -3.82$\pm$0.08 & -4.34$\pm$0.12 & 2775.3$\pm$1.6 & 432.8$\pm$0.4 \\
\hline
         &             &             & 0.65 &  0.41$\pm$0.17 &  0.72$\pm$0.10 & -0.80$\pm$0.17 & -1.18$\pm$0.10 & 654.0$\pm$0.8 & 644.3$\pm$0.4 \\
Source 1 & 09:55:47.29 & 69:40:48.37 & 0.70 &  0.42$\pm$0.17 &  0.64$\pm$0.10 & -1.03$\pm$0.16 & -1.28$\pm$0.10 & 686.8$\pm$0.8 & 676.2$\pm$0.5 \\
         &             &             & 0.75 &  0.46$\pm$0.16 &  0.55$\pm$0.09 & -1.13$\pm$0.16 & -1.45$\pm$0.09 & 720.6$\pm$0.8 & 707.6$\pm$0.5 \\
\hline
         &             &             & 0.65 &  3.55$\pm$0.06 &  3.60$\pm$0.03 & -2.98$\pm$0.06 & -3.29$\pm$0.03 & 5362.3$\pm$2.2 & 5373.5$\pm$1.3 \\
Source 2 & 09:55:46.97 & 69:40:41.73 & 0.70 &  3.50$\pm$0.06 &  3.53$\pm$0.03 & -2.95$\pm$0.06 & -3.24$\pm$0.03 & 5573.6$\pm$2.3 & 5583.6$\pm$1.3 \\
         &             &             & 0.75 &  3.43$\pm$0.06 &  3.46$\pm$0.03 & -2.89$\pm$0.06 & -3.23$\pm$0.03 & 5779.5$\pm$2.3 & 5787.4$\pm$1.3 \\
\hline
         &             &             & 0.35 &  1.18$\pm$0.26 &  1.33$\pm$0.15 & -2.64$\pm$0.26 & -2.81$\pm$0.15 & 272.5$\pm$0.5 & 274.8$\pm$0.3 \\
Source 3 & 09:55:46.51 & 69:40:43.37 & 0.40 &  1.39$\pm$0.24 &  1.38$\pm$0.14 & -2.52$\pm$0.24 & -2.85$\pm$0.13 & 325.4$\pm$0.5 & 329.8$\pm$0.3 \\
         &             &             & 0.45 &  1.15$\pm$0.23 &  1.39$\pm$0.13 & -2.22$\pm$0.22 & -2.96$\pm$0.13 & 370.6$\pm$0.6 & 375.2$\pm$0.3 \\
\hline
         &             &             & 0.35 &  0.38$\pm$0.23 & -3.41$\pm$0.23 &  0.86$\pm$0.13 & -3.95$\pm$0.13 &  338.4$\pm$0.5 &  334.2$\pm$0.3 \\
Source 4 & 09:55:43.95 & 69:40:35.49 & 0.40 &  0.50$\pm$0.22 & -3.32$\pm$0.21 &  0.54$\pm$0.12 & -3.86$\pm$0.12 &  391.0$\pm$0.6 &  387.1$\pm$0.3 \\
         &             &             & 0.45 &  0.61$\pm$0.20 & -3.25$\pm$0.20 &  0.42$\pm$0.12 & -3.67$\pm$0.12 &  437.8$\pm$0.6 &  433.5$\pm$0.4 \\
\hline
\end{tabular}
{$^1$}{Measurement of $F475W$ from epoch 1 at t=277 days. \\
$^2$Measurement of $F475W$ from epoch 2 at t=416 days.} 
\end{scriptsize}
\end{table}


\end{document}